\documentclass[zpreprint,zbstpl]{./zeus_paper}
\usepackage[english]{babel}
\usepackage{lineno}
\chardef\usc=95
\chardef\til=126
\catcode`\@=11 
\DeclareRobustCommand\xdotspace{\futurelet\@let@token\@xdotspace}
\def\@xdotspace{%
  \ifx\@let@token.\else
  \ifx\@let@token\bgroup.\else
  \ifx\@let@token\egroup.\else
  \ifx\@let@token\/.\else
  \ifx\@let@token\ .\else
  \ifx\@let@token~.\else
  \ifx\@let@token!.\else
  \ifx\@let@token,.\else
  \ifx\@let@token:.\else
  \ifx\@let@token;.\else
  \ifx\@let@token?.\else
  \ifx\@let@token/.\else
  \ifx\@let@token'.\else
  \ifx\@let@token).\else
  \ifx\@let@token-.\else
  \ifx\@let@token\@xobeysp.\else
  \ifx\@let@token\space.\else
  \ifx\@let@token\@sptoken.\else
   .\space
   \fi\fi\fi\fi\fi\fi\fi\fi\fi\fi\fi\fi\fi\fi\fi\fi\fi\fi}
\catcode`\@=12 

\newcommand{\stru}[2]{%
   \relax\ifmmode\hbox{\vrule height#1 depth#2 width0pt}%
   \else\vrule height#1 depth#2 width0pt\fi}

\newcommand{\Ronum}[1]{\uppercase\expandafter{\romannumeral#1}}
\newcommand{\ronum}[1]{\expandafter{\romannumeral#1}}
\DeclareRobustCommand{\LaTeXZ}{%
  \LaTeX\kern-.05em4\kern-.1em
  {\raisebox{-0.2ex}{$\scriptstyle\text{ZEUS}$}}\xspace}

\DeclareMathAlphabet{\mathbf}{OT1}{cmr}{bx}{sl}
\newcommand{\eVdist}{\kern-0.06667em}

\newcommand{\Gev}{{\text{Ge}\eVdist\text{V\/}}}

\newcommand{\gev}{{\,\text{Ge}\eVdist\text{V\/}}}

\newcommand{\met}{\,\text{m}}
\newcommand{\nm}{\,\text{nm}}

\newcommand{\mm}{\,\text{mm}}
\newcommand{\cm}{\,\text{cm}}
\newcommand{\km}{\,\text{km}}

\newcommand{\Hz}{\,\text{Hz}}
\newcommand{\kHz}{\,\text{kHz}}
\newcommand{\MHz}{\,\text{MHz}}
\newcommand{\GHz}{\,\text{GHz}}

\newcommand{\rad}{\,\text{rad}}

\newcommand{\mrad}{\,\text{mrad}}

\newcommand{\slashfrac}[2]{%
  \raisebox{0.5ex}{\ensuremath #1}\kern-0.12em/\kern-0.08em
  \raisebox{-.8ex}{\ensuremath #2}}

\newcommand{\sqr}[3]{%
    {\vcenter{\hrule height.#3ex\hbox{\vrule width.#2ex height#1ex
     \kern#1ex\vrule width.#3ex}\hrule height.#2ex}}}

\catcode`\@=11 
\newcommand{\parenbar}{\mathpalette\p@renb@r}
\def\p@renb@r#1#2{\vbox{%
  \ifx#1\scriptscriptstyle \dimen@.7em\dimen@ii.2em\else
  \ifx#1\scriptstyle \dimen@.8em\dimen@ii.25em\else
  \dimen@1em\dimen@ii.4em\fi\fi \offinterlineskip
  \ialign{\hfill##\hfill\cr
    \vbox{\hrule width\dimen@ii}\cr
    \noalign{\vskip-.3ex}%
    \hbox to\dimen@{$\mathchar300\hfil\mathchar301$}\cr
    \noalign{\vskip-.3ex}%
    $#1#2$\cr}}}
\catcode`\@=12 

\newcommand{\IP}{{\rm I$\kern-0.01667em$P}\xspace}
\newcommand{\JB}{{\rm JB}}

\newcommand{\jet}{{\rm jet}}

\mathchardef\qsm=63
\mathchardef\pls=43
\mathchardef\mns=512
\mathchardef\plm=518
\mathchardef\eql=61
\mathchardef\smallleft=300
\mathchardef\smallright=301
\mathchardef\les=316
\mathchardef\gre=318
\mathchardef\leq=532
\mathchardef\grq=533

\catcode`\@=11 
\newcounter{pict@width}
\newcounter{pict@height}
\newlength{\pict@scale}
\newcommand{\psfigadd}[4]{%
\setcounter{pict@width}{1*\ratio{#2+\pict@scale/2}{\pict@scale}}
\setcounter{pict@height}{1*\ratio{#3+\pict@scale/2}{\pict@scale}}
\setlength{\unitlength}{\pict@scale}
\hbox to #2{\hspace{-\fill}\begin{picture}(\thepict@width,\thepict@height)
\put(0,0){\psfig{figure=#1,width=#2,height=#3,clip=}}
\SetScale{0.283466457}
\SetWidth{1.763889}
{#4}
\end{picture}}
}
\newcounter{pict@widthfst}
\newcounter{pict@widthscd}
\newcounter{pict@widthtot}
\newcommand{\psfigaddtwo}[7]{%
\setcounter{pict@widthfst}{1*\ratio{#2+\pict@scale/2}{\pict@scale}}
\setcounter{pict@widthscd}{1*\ratio{#2+#4+\pict@scale/2}{\pict@scale}}
\setcounter{pict@widthtot}{1*\ratio{#2+#4+#6+\pict@scale/2}{\pict@scale}}
\setcounter{pict@height}{1*\ratio{#3+\pict@scale/2}{\pict@scale}}
\setlength{\unitlength}{\pict@scale}
\hbox{\hspace{-\fill}\begin{picture}(\thepict@widthtot,\thepict@height)
\put(0,0){\psfig{figure=#1,width=#2,height=#3,clip=}}
\put(\thepict@widthscd,0){\psfig{figure=#5,width=#6,height=#3,clip=}}
\SetScale{0.283466457}
\SetWidth{1.763889}
{#7}
\end{picture}}
}
\newcommand{\psfigror}[4]{%
\setcounter{pict@width}{1*\ratio{#2+\pict@scale/2}{\pict@scale}}
\setcounter{pict@height}{1*\ratio{#3+\pict@scale/2}{\pict@scale}}
\setlength{\unitlength}{\pict@scale}
\hbox{\begin{picture}(\thepict@width,\thepict@height)
\put(0,\thepict@height){\psfig{figure=#1,width=#3,height=#2,clip=,angle=270}}
\SetScale{0.283466457}
\SetWidth{1.763889}
{#4}
\end{picture}}
}
\newcommand{\psfigrol}[4]{%
\setcounter{pict@width}{1*\ratio{#2+\pict@scale/2}{\pict@scale}}
\setcounter{pict@height}{1*\ratio{#3+\pict@scale/2}{\pict@scale}}
\setlength{\unitlength}{\pict@scale}
\hbox{\begin{picture}(\thepict@width,\thepict@height)
\put(0,0){\psfig{figure=#1,width=#3,height=#2,clip=,angle=90}}
\SetScale{0.283466457}
\SetWidth{1.763889}
{#4}
\end{picture}}
}
\catcode`\@=12 
\newlength\listtextwidth

\catcode`\@=11 
\newlength{\@tabfninsert}
\newlength{\@tabfnwidth}
\newcommand{\tabfootnote}[2]{%
  \setlength{\@tabfninsert}{0.8em}
  \setlength{\@tabfnwidth}{\textwidth}
  \addtolength{\@tabfnwidth}{-\@tabfninsert}
  \addtolength{\@tabfnwidth}{-0.4em}
  \noindent\makebox[\@tabfninsert][r]{\footnotesize$^{#1}$\hfil}\hfill%
  \parbox[t]{\@tabfnwidth}{\footnotesize #2\hfill}}
\catcode`\@=12 

\newcommand{\ZcoosysAA}{%
The ZEUS coordinate system is a right-handed Cartesian system, with the $Z$
axis pointing in the proton beam direction, referred to as the ``forward
direction'', and the $X$ axis pointing  towards the center of HERA.
The coordinate origin is at the nominal interaction point.\xspace}
\newcommand{\ZcoosysAE}{%
The ZEUS coordinate system is a right-handed Cartesian system, with the $Z$
axis pointing in the proton beam direction, referred to as the ``forward
direction'', and the $X$ axis pointing  towards the centre of HERA.
The coordinate origin is at the nominal interaction point.\xspace}
\newcommand{\ZcoosysBA}{%
The ZEUS coordinate system is a right-handed Cartesian system, with the $Z$ 
axis pointing in the nominal proton beam direction, referred to as the ``forward
direction'', and the $X$ axis pointing  towards the center of HERA.
The coordinate origin is at the centre of the CTD.\xspace}
\newcommand{\ZcoosysBE}{%
The ZEUS coordinate system is a right-handed Cartesian system, with the $Z$ 
axis pointing in the nominal proton beam direction, referred to as the ``forward
direction'', and the $X$ axis pointing  towards the centre of HERA.
The coordinate origin is at the centre of the CTD.\xspace}
\newcommand{\ZpsrapA}{%
The pseudorapidity is defined as $\eta=-\ln\left(\tan\frac{\theta}{2}\right)$,
where the polar angle, $\theta$, is measured with respect to the
$Z$ axis.\xspace}
\newcommand{\ZpsrapB}{%
The pseudorapidity is defined as $\eta=-\ln\left(\tan\frac{\theta}{2}\right)$, 
where the polar angle, $\theta$, is measured with respect to the
$Z$ axis. The azimuthal angle, $\phi$, is
measured with respect to the $X$ axis.\xspace}

\newcommand{\ZcoosysfnBEeta}{\footnote{\ZcoosysBE\ZpsrapA}}

\newcommand{\Zsttdesc}[1]{%
The STT consisted of 48 sectors of two different sizes. Each sector
contained 192 (small sector) or 264 (large sector) straws of diameter
\unit[7.5]{\mm} arranged into 3 layers. The sectors were trapezoidal in shape
and each subtended an azimuthal angle of $60\degree$ -- 6 sectors
formed a so-called superlayer. A particle passing through the complete
detector traversed 8 superlayers, which were rotated around the beam
direction at angles of $30\degree$ or $15\degree$ to each other. The STT
covered the polar-angle region $5\degree < \theta < 23\degree$.
}
\def\citeCTD{{\cite{%
nim:a279:290,*npps:b32:181,*nim:a338:254%
}}\xspace}
\def\citeMVD{{\cite{%
nim:a581:656%
}}\xspace}
\def\citeCAL{{\cite{%
nim:a309:77,*nim:a309:101,*nim:a321:356,*nim:a336:23%
}}\xspace}

\newcommand{\PYTHIA}{\textsc{Pythia}\xspace}
\newcommand{\HERWIG}{\textsc{Herwig}\xspace}

\newcommand{\ET}{\ensuremath{E_{T}}\xspace}
\newcommand{\ETjet}{\ensuremath{E_{T}^{\mathrm{jet}}}\xspace}
\newcommand{\Ejet}{\ensuremath{E^{\mathrm{jet}}}\xspace}
\newcommand{\pzjet}{\ensuremath{p_Z^{\mathrm{jet}}}\xspace}

\newcommand{\Egam}{\ensuremath{E^{\gamma}}\xspace}
\newcommand{\pzgam}{\ensuremath{p_Z^{\gamma}}\xspace}
\newcommand{\ETgam}{\ensuremath{E_{T}^{\gamma}}\xspace}
\newcommand{\etagam}{\ensuremath{\eta^{\gamma}}\xspace}
\newcommand{\etajet}{\ensuremath{\eta^{\mathrm{jet}}}\xspace}

\newcommand{\xgamm}{\ensuremath{x_{\gamma}^{\mathrm{meas}}}\xspace}
\newcommand{\Zacknowledge}{%
We appreciate the contributions to the construction, maintenance and operation of
the ZEUS detector made by many people who are not listed as authors. The
HERA machine group and the DESY computing staff are especially
acknowledged for their success in providing excellent operation of the
collider and the data-analysis environment. We thank the DESY
directorate for their strong support and encouragement.}

\begin{document}
\prepnum{DESY-13-234}

\title{
Photoproduction of isolated photons, inclusively and with a jet, at HERA
}

\author{ZEUS Collaboration}
\date{\today}

\abstract{
The photoproduction of isolated photons, both inclusive and
together with a jet, has been measured with the ZEUS detector at HERA
using an integrated luminosity of $374\, \mathrm{pb}^{-1}$.
Differential cross sections are presented in the isolated-photon
transverse-energy and pseudorapidity ranges \mbox{$6 < E_T^{\gamma} <
15\,\gev$} and \mbox{$-0.7 < \eta^{\gamma} < 0.9$,} and for jet
transverse-energy and pseudorapidity ranges $4 < \ETjet <
35\,\gev$ and $-1.5 < \etajet < 1.8$, for exchanged-photon
virtualities $Q^2 < 1 \gev^2$.  Differential cross sections
are also presented for inclusive isolated-photon production as
functions of the transverse energy and pseudorapidity of the
photon. Higher-order theoretical calculations are compared to the
results. }

\makezeustitle

%
%
%
%


\def\3{\ss}
\newcommand{\address}{ }
\newcommand{\author}{ }
\pagenumbering{Roman}
                                                   %
\begin{center}
{                      \Large  The ZEUS Collaboration              }
\end{center}

{\small


        {\raggedright
H.~Abramowicz$^{27, u}$, 
I.~Abt$^{21}$, 
L.~Adamczyk$^{8}$, 
M.~Adamus$^{34}$, 
R.~Aggarwal$^{4, a}$, 
S.~Antonelli$^{2}$, 
O.~Arslan$^{3}$, 
V.~Aushev$^{16, 17, o}$, 
Y.~Aushev,$^{17, o, p}$, 
O.~Bachynska$^{10}$
% ...
        }

\newpage


\makebox[3em]{$^{1}$}
\begin{minipage}[t]{14cm}
{\it INFN Bologna, Bologna, Italy}~$^{A}$

\end{minipage}\\
\makebox[3em]{$^{2}$}
\begin{minipage}[t]{14cm}
{\it University and INFN Bologna, Bologna, Italy}~$^{A}$

\end{minipage}\\
\makebox[3em]{$^{3}$}
\begin{minipage}[t]{14cm}
{\it Physikalisches Institut der Universit\"at Bonn,
Bonn, Germany}~$^{B}$

\end{minipage}\\
\makebox[3em]{$^{4}$}
\begin{minipage}[t]{14cm}
{\it Panjab University, Department of Physics, Chandigarh, India}

\end{minipage}\\
\makebox[3em]{$^{5}$}
\begin{minipage}[t]{14cm}
{\it Calabria University,
Physics Department and INFN, Cosenza, Italy}~$^{A}$

\end{minipage}\\
\makebox[3em]{$^{6}$}
\begin{minipage}[t]{14cm}
{\it National Centre for Particle Physics, Universiti Malaya, 50603 Kuala Lumpur, Malaysia}~$^{C}$

\end{minipage}\\
\makebox[3em]{$^{7}$}
\begin{minipage}[t]{14cm}
{\it The Henryk Niewodniczanski Institute of Nuclear Physics, Polish Academy of \\
Sciences, Krakow, Poland}~$^{D}$

\end{minipage}\\
\makebox[3em]{$^{8}$}
\begin{minipage}[t]{14cm}
{\it AGH-University of Science and Technology, Faculty of Physics and Applied Computer
Science, Krakow, Poland}~$^{D}$

\end{minipage}\\
\makebox[3em]{$^{9}$}
\begin{minipage}[t]{14cm}
{\it Department of Physics, Jagellonian University, Cracow, Poland}

\end{minipage}\\
\makebox[3em]{$^{10}$}
\begin{minipage}[t]{14cm}
{\it Deutsches Elektronen-Synchrotron DESY, Hamburg, Germany}

\end{minipage}\\
\makebox[3em]{$^{11}$}
\begin{minipage}[t]{14cm}
{\it Deutsches Elektronen-Synchrotron DESY, Zeuthen, Germany}

\end{minipage}\\
\makebox[3em]{$^{12}$}
\begin{minipage}[t]{14cm}
{\it School of Physics and Astronomy, University of Glasgow,
Glasgow, United Kingdom}~$^{E}$

\end{minipage}\\
\makebox[3em]{$^{13}$}
\begin{minipage}[t]{14cm}
{\it Hamburg University, Institute of Experimental Physics, Hamburg,
Germany}~$^{F}$

\end{minipage}\\
\makebox[3em]{$^{14}$}
\begin{minipage}[t]{14cm}
{\it Institute of Particle and Nuclear Studies, KEK,
Tsukuba, Japan}~$^{G}$

\end{minipage}\\
\makebox[3em]{$^{15}$}
\begin{minipage}[t]{14cm}
{\it Institute of Physics and Technology of Ministry of Education and
Science of Kazakhstan, Almaty, Kazakhstan}

\end{minipage}\\
\makebox[3em]{$^{16}$}
\begin{minipage}[t]{14cm}
{\it Institute for Nuclear Research, National Academy of Sciences, Kyiv, Ukraine}

\end{minipage}\\
\makebox[3em]{$^{17}$}
\begin{minipage}[t]{14cm}
{\it Department of Nuclear Physics, National Taras Shevchenko University of Kyiv, Kyiv, Ukraine}

\end{minipage}\\
\makebox[3em]{$^{18}$}
\begin{minipage}[t]{14cm}
{\it Department of Physics, McGill University,
Montr\'eal, Qu\'ebec, Canada H3A 2T8}~$^{H}$

\end{minipage}\\
\makebox[3em]{$^{19}$}
\begin{minipage}[t]{14cm}
{\it Meiji Gakuin University, Faculty of General Education,
Yokohama, Japan}~$^{G}$

\end{minipage}\\
\makebox[3em]{$^{20}$}
\begin{minipage}[t]{14cm}
{\it Lomonosov Moscow State University, Skobeltsyn Institute of Nuclear Physics,
Moscow, Russia}~$^{I}$

\end{minipage}\\
\makebox[3em]{$^{21}$}
\begin{minipage}[t]{14cm}
{\it Max-Planck-Institut f\"ur Physik, M\"unchen, Germany}

\end{minipage}\\
\makebox[3em]{$^{22}$}
\begin{minipage}[t]{14cm}
{\it Department of Physics, University of Oxford,
Oxford, United Kingdom}~$^{E}$

\end{minipage}\\
\makebox[3em]{$^{23}$}
\begin{minipage}[t]{14cm}
{\it INFN Padova, Padova, Italy}~$^{A}$

\end{minipage}\\
\makebox[3em]{$^{24}$}
\begin{minipage}[t]{14cm}
{\it Dipartimento di Fisica dell' Universit\`a and INFN,
Padova, Italy}~$^{A}$

\end{minipage}\\
\makebox[3em]{$^{25}$}
\begin{minipage}[t]{14cm}
{\it Polytechnic University, Tokyo, Japan}~$^{G}$

\end{minipage}\\
\makebox[3em]{$^{26}$}
\begin{minipage}[t]{14cm}
{\it Dipartimento di Fisica, Universit\`a `La Sapienza' and INFN,
Rome, Italy}~$^{A}$

\end{minipage}\\
\makebox[3em]{$^{27}$}
\begin{minipage}[t]{14cm}
{\it Raymond and Beverly Sackler Faculty of Exact Sciences, School of Physics, \\
Tel Aviv University, Tel Aviv, Israel}~$^{J}$

\end{minipage}\\
\makebox[3em]{$^{28}$}
\begin{minipage}[t]{14cm}
{\it Department of Physics, Tokyo Institute of Technology,
Tokyo, Japan}~$^{G}$

\end{minipage}\\
\makebox[3em]{$^{29}$}
\begin{minipage}[t]{14cm}
{\it Universit\`a di Torino and INFN, Torino, Italy}~$^{A}$

\end{minipage}\\
\makebox[3em]{$^{30}$}
\begin{minipage}[t]{14cm}
{\it Universit\`a del Piemonte Orientale, Novara, and INFN, Torino,
Italy}~$^{A}$

\end{minipage}\\
\makebox[3em]{$^{31}$}
\begin{minipage}[t]{14cm}
{\it Department of Physics, University of Toronto, Toronto, Ontario,
Canada M5S 1A7}~$^{H}$

\end{minipage}\\
\makebox[3em]{$^{32}$}
\begin{minipage}[t]{14cm}
{\it Physics and Astronomy Department, University College London,
London, United Kingdom}~$^{E}$

\end{minipage}\\
\makebox[3em]{$^{33}$}
\begin{minipage}[t]{14cm}
{\it Faculty of Physics, University of Warsaw, Warsaw, Poland}

\end{minipage}\\
\makebox[3em]{$^{34}$}
\begin{minipage}[t]{14cm}
{\it National Centre for Nuclear Research, Warsaw, Poland}

\end{minipage}\\
\makebox[3em]{$^{35}$}
\begin{minipage}[t]{14cm}
{\it Department of Particle Physics and Astrophysics, Weizmann
Institute, Rehovot, Israel}

\end{minipage}\\
\makebox[3em]{$^{36}$}
\begin{minipage}[t]{14cm}
{\it Department of Physics, York University, Ontario, Canada M3J 1P3}~$^{H}$

\end{minipage}\\
\vspace{30em} \pagebreak[4]


\makebox[3ex]{$^{ A}$}
\begin{minipage}[t]{14cm}
 supported by the Italian National Institute for Nuclear Physics (INFN) \
\end{minipage}\\
\makebox[3ex]{$^{ B}$}
\begin{minipage}[t]{14cm}
 supported by the German Federal Ministry for Education and Research (BMBF), under
 contract No. 05 H09PDF\
\end{minipage}\\
\makebox[3ex]{$^{ C}$}
\begin{minipage}[t]{14cm}
 supported by HIR grant UM.C/625/1/HIR/149 and UMRG grants RU006-2013, RP012A-13AFR and RP012B-13AFR from
 Universiti Malaya, and ERGS grant ER004-2012A from the Ministry of Education, Malaysia\
\end{minipage}\\
\makebox[3ex]{$^{ D}$}
\begin{minipage}[t]{14cm}
 supported by the National Science Centre under contract No. DEC-2012/06/M/ST2/00428\
\end{minipage}\\
\makebox[3ex]{$^{ E}$}
\begin{minipage}[t]{14cm}
 supported by the Science and Technology Facilities Council, UK\
\end{minipage}\\
\makebox[3ex]{$^{ F}$}
\begin{minipage}[t]{14cm}
 supported by the German Federal Ministry for Education and Research (BMBF), under
 contract No. 05h09GUF, and the SFB 676 of the Deutsche Forschungsgemeinschaft (DFG) \
\end{minipage}\\
\makebox[3ex]{$^{ G}$}
\begin{minipage}[t]{14cm}
 supported by the Japanese Ministry of Education, Culture, Sports, Science and Technology
 (MEXT) and its grants for Scientific Research\
\end{minipage}\\
\makebox[3ex]{$^{ H}$}
\begin{minipage}[t]{14cm}
 supported by the Natural Sciences and Engineering Research Council of Canada (NSERC) \
\end{minipage}\\
\makebox[3ex]{$^{ I}$}
\begin{minipage}[t]{14cm}
 supported by RF Presidential grant N 3920.2012.2 for the Leading Scientific Schools and by
 the Russian Ministry of Education and Science through its grant for Scientific Research on
 High Energy Physics\
\end{minipage}\\
\makebox[3ex]{$^{ J}$}
\begin{minipage}[t]{14cm}
 supported by the Israel Science Foundation\
\end{minipage}\\
\vspace{30em} \pagebreak[4]


\makebox[3ex]{$^{ a}$}
\begin{minipage}[t]{14cm}
also funded by Max Planck Institute for Physics, Munich, Germany\
\end{minipage}\\
\makebox[3ex]{$^{ b}$}
\begin{minipage}[t]{14cm}
also funded by Max Planck Institute for Physics, Munich, Germany, now at Sri Guru Granth Sahib World University, Fatehgarh Sahib\
\end{minipage}\\
\makebox[3ex]{$^{ c}$}
\begin{minipage}[t]{14cm}
also at Agensi Nuklear Malaysia, 43000 Kajang, Bangi, Malaysia\
\end{minipage}\\
\makebox[3ex]{$^{ d}$}
\begin{minipage}[t]{14cm}
also at Cracow University of Technology, Faculty of Physics, Mathematics and Applied Computer Science, Poland\
\end{minipage}\\
\makebox[3ex]{$^{ e}$}
\begin{minipage}[t]{14cm}
partially supported by the Polish National Science Centre projects DEC-2011/01/B/ST2/03643 and DEC-2011/03/B/ST2/00220\
\end{minipage}\\
\makebox[3ex]{$^{ f}$}
\begin{minipage}[t]{14cm}
now at Rockefeller University, New York, NY 10065, USA\
\end{minipage}\\
\makebox[3ex]{$^{ g}$}
\begin{minipage}[t]{14cm}
now at University of Liverpool, United Kingdom\
\end{minipage}\\
\makebox[3ex]{$^{ h}$}
\begin{minipage}[t]{14cm}
also affiliated with University College London, UK\
\end{minipage}\\
\makebox[3ex]{$^{ i}$}
\begin{minipage}[t]{14cm}
supported by the Alexander von Humboldt Foundation\
\end{minipage}\\
\makebox[3ex]{$^{ j}$}
\begin{minipage}[t]{14cm}
now at CERN, Geneva, Switzerland\
\end{minipage}\\
\makebox[3ex]{$^{ k}$}
\begin{minipage}[t]{14cm}
also at Institute of Theoretical and Experimental Physics, Moscow, Russia\
\end{minipage}\\
\makebox[3ex]{$^{ l}$}
\begin{minipage}[t]{14cm}
Alexander von Humboldt Professor; also at DESY and University of Oxford\
\end{minipage}\\
\makebox[3ex]{$^{ m}$}
\begin{minipage}[t]{14cm}
also at University of Tokyo, Japan\
\end{minipage}\\
\makebox[3ex]{$^{ n}$}
\begin{minipage}[t]{14cm}
now at Kobe University, Japan\
\end{minipage}\\
\makebox[3ex]{$^{ o}$}
\begin{minipage}[t]{14cm}
supported by DESY, Germany\
\end{minipage}\\
\makebox[3ex]{$^{ p}$}
\begin{minipage}[t]{14cm}
member of National Technical University of Ukraine, Kyiv Polytechnic Institute, Kyiv, Ukraine\
\end{minipage}\\
\makebox[3ex]{$^{ q}$}
\begin{minipage}[t]{14cm}
now at DESY ATLAS group\
\end{minipage}\\
\makebox[3ex]{$^{ r}$}
\begin{minipage}[t]{14cm}
member of National University of Kyiv - Mohyla Academy, Kyiv, Ukraine\
\end{minipage}\\
\makebox[3ex]{$^{ s}$}
\begin{minipage}[t]{14cm}
now at BNL, USA\
\end{minipage}\\
\makebox[3ex]{$^{ t}$}
\begin{minipage}[t]{14cm}
now at LNF, Frascati, Italy\
\end{minipage}\\
\makebox[3ex]{$^{ u}$}
\begin{minipage}[t]{14cm}
also at Max Planck Institute for Physics, Munich, Germany, External Scientific Member\
\end{minipage}\\
\makebox[3ex]{$^{ v}$}
\begin{minipage}[t]{14cm}
now at Tokyo Metropolitan University, Japan\
\end{minipage}\\
\makebox[3ex]{$^{ w}$}
\begin{minipage}[t]{14cm}
also supported by DESY\
\end{minipage}\\
\makebox[3ex]{$^{ x}$}
\begin{minipage}[t]{14cm}
also at \L\'{o}d\'{z} University, Poland\
\end{minipage}\\
\makebox[3ex]{$^{ y}$}
\begin{minipage}[t]{14cm}
member of \L\'{o}d\'{z} University, Poland\
\end{minipage}\\
\makebox[3ex]{$^{ z}$}
\begin{minipage}[t]{14cm}
now at Polish Air Force Academy in Deblin\
\end{minipage}\\

}


\pagenumbering{arabic} 
\pagestyle{plain}

\section{Introduction}
\label{sec-int}

Events containing an isolated high-energy photon can provide
a direct probe of the underlying partonic process in high-energy
collisions involving hadrons, since the emission of such photons is
largely unaffected by parton hadronisation. 
Processes of this kind have
been studied in a number of fixed-target and hadron-collider
experiments
\cite{pr:d73:094007,*prl:95:022003,*pl:b639:151,%
*D0pre13,*np:b875:483,*ATLASnew13,*pr:d84:052011,*pl:b710:403}.
In $ep$ collisions at HERA, the ZEUS and H1 collaborations have
previously reported the production of isolated photons in
photoproduction
\cite{pl:b413:201,pl:b472:175,pl:b511:19,epj:c49:511,epj:c38:437,epj:c66:17}, 
in which the exchanged virtual photon is quasi-real, and also in deep
inelastic scattering
(DIS)~\cite{pl:b595:86,epj:c54:371,pl:b687:16,pl:b715:88}.  In this
paper, earlier photoproduction measurements by ZEUS are extended by
using the full HERA II data set. The statistical precision is
much improved owing to the availability of higher integrated
luminosity.  Measurements are presented of isolated-photon production
at high transverse energy with and without an explicit accompanying-jet
requirement.  The measurement of the jet gives further information on the
event dynamics. 

\begin{figure}
\vfill
\begin{center}
\epsfig{file=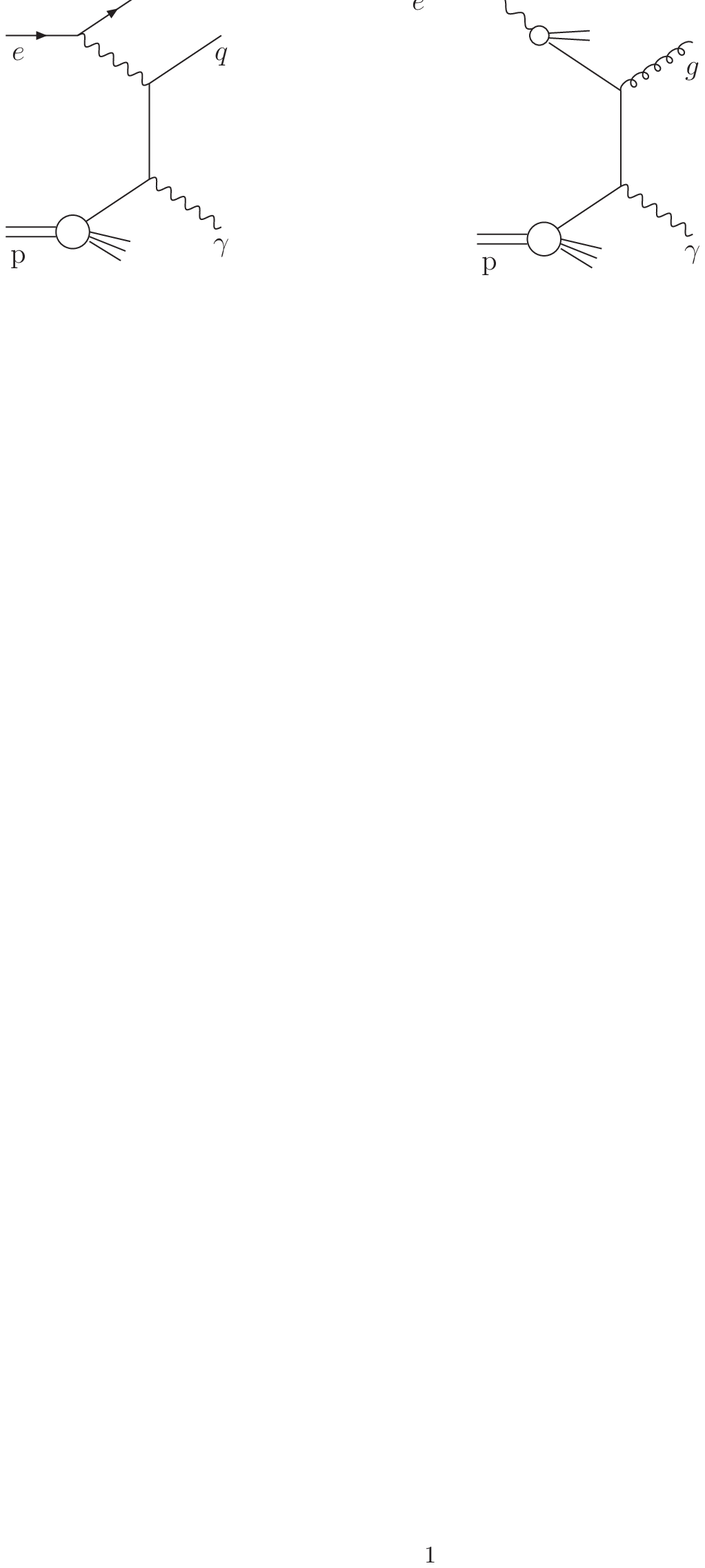,width=11cm,%
bbllx=120,bblly=670,bburx=420,bbury=820}
\hspace*{10mm}
\\[-4mm]
(a)\hspace{7.5cm}(b)\\[5mm]
\epsfig{file=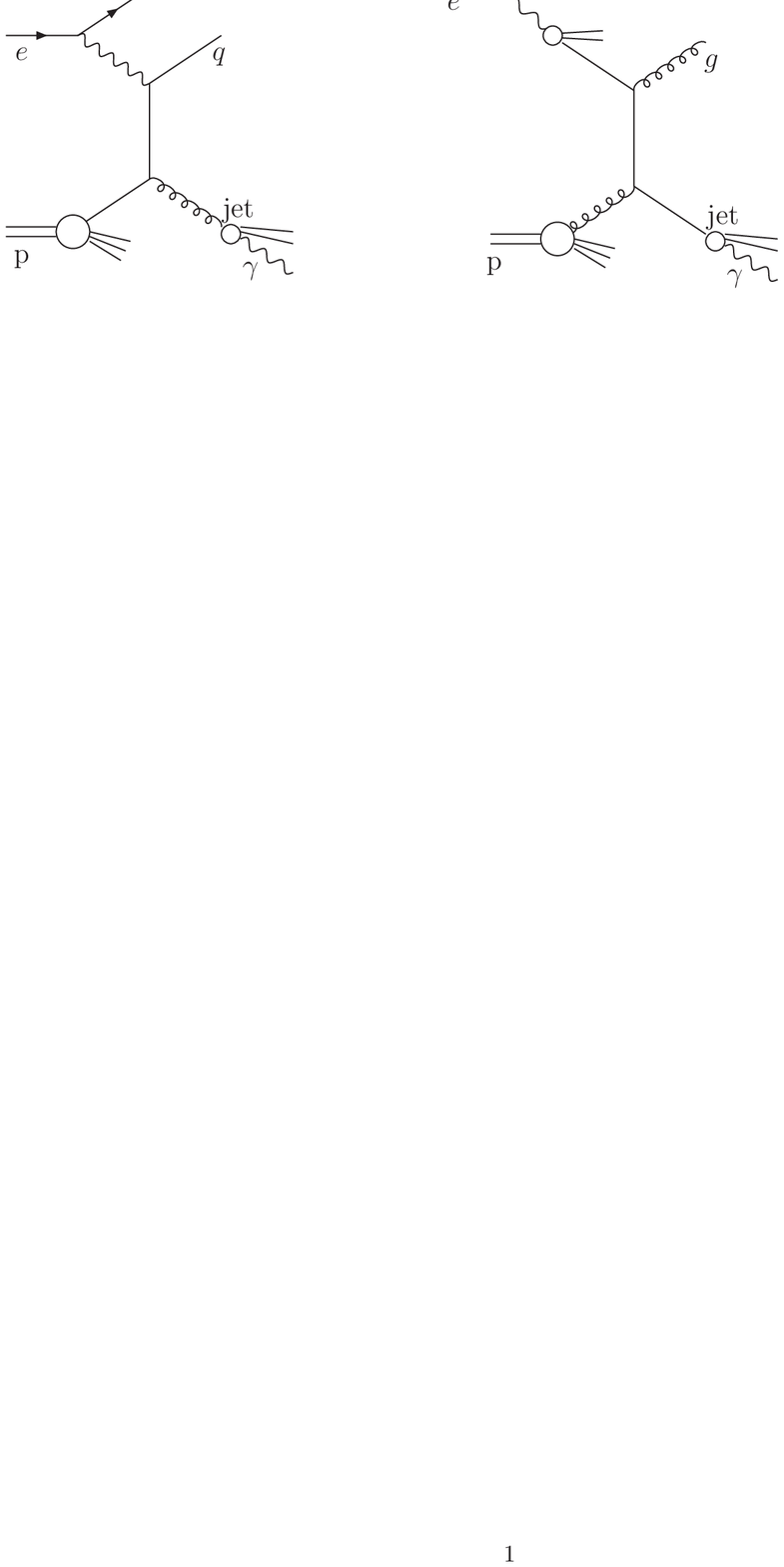,width=12cm,%
bbllx=90,bblly=670,bburx=420,bbury=820}
\\[-4mm]
(c)\hspace{7.5cm}(d)
\end{center}
\vspace*{0mm}
\caption{Examples of (a) direct-prompt and (b) resolved-prompt 
processes at leading order in QCD, and the related (c) direct and 
(d) resolved fragmentation processes.}
\label{fig1}
\vfill
\end{figure}

Figure~\ref{fig1} gives examples of the lowest-order (LO) diagrams for
high-energy photoproduction of photons in quantum chromodynamics (QCD).
In ``direct'' production processes, the entire incoming photon is
absorbed by a quark from the incoming proton, while in ``resolved''
processes, the photon's hadronic structure provides a quark or gluon
that interacts with a parton from the proton. Photons that are
radiated in the hard scattering process, rather than resulting from
meson decay, are commonly called ``prompt''\footnote{An alternative
commonly-used nomenclature is to refer to ``prompt'' photons as
``direct''; thus Figs.~\ref{fig1}(a) and
\ref{fig1}(b) would be called ``direct-direct'' and ``resolved-direct''
diagrams, respectively.}.  Higher-order processes include 
``fragmentation processes'' in which a photon is radiated
within a jet, also illustrated in Fig.~\ref{fig1}. Such processes are
suppressed by requiring that the photon be isolated.  Photons radiated
at large angles from the incoming or outgoing electron give rise to an
observed scattered electron in the detector; such events are
excluded from this measurement.

Perturbative QCD predictions are compared to the measurements.  The
cross sections for isolated-photon production in photoproduction have
been calculated to next-to-leading order (NLO) by Fontannaz,
Guillet and Heinrich (FGH)~\cite{epj:c21:303,epj:c34:191}.  
Calculations based on the $k_T$-factorisation approach have been made by
Lipatov, Malyshev and Zotov 
(LMZ)~\cite{pr:d72:054002,pr:d81:094027,pr:d88:074001}.

\section{Experimental set-up}
\label{sec-exp}
The measurements are based on a data sample corresponding to an
integrated luminosity of $374\pm 7 \,\mathrm{pb}^{-1}$, taken during the
years 2004 to 2007 with the ZEUS detector at
HERA. During this period, HERA ran with an electron or positron beam
energy of 27.5\gev\ and a proton beam energy of 920\gev. The sample
is a sum of $e^+p$ and $e^-p$ data\footnote{Hereafter ``electron'' refers to both electrons
and positrons unless otherwise stated.}.

A detailed description of the ZEUS detector can be found
elsewhere~\cite{zeus:1993:bluebook}. Charged particles were measured in
the central tracking detector (CTD)~\citeCTD and a silicon micro
vertex detector (MVD)~\cite{nim:a581:656} which operated in a magnetic
field of $1.43$~T provided by a thin superconducting solenoid.  The
high-resolution uranium--scintillator calorimeter (CAL)~\citeCAL
consisted of three parts: the forward (FCAL), the barrel (BCAL) and
the rear (RCAL) calorimeters. The BCAL covered the pseudorapidity
range --0.74 to 1.01 as seen from the nominal interaction point, and the
FCAL and RCAL extended the coverage to the range --3.5 to 4.0.  Each
part of the CAL was subdivided into elements referred to as cells. The
barrel electromagnetic calorimeter (BEMC) cells had a pointing
geometry aimed at the nominal interaction point, with a cross section
approximately $5\times20\,\mathrm{cm^2}$, with the finer granularity
in the $Z$ direction{\ZcoosysfnBEeta} and the coarser in the $(X,Y)$
plane.  This fine granularity allows the use of shower-shape
distributions to distinguish isolated photons from the products of
neutral meson decays such as $\pi^0 \rightarrow \gamma\gamma$.
The CAL energy resolution, as measured under test-beam conditions, was
$\sigma(E)/E = 0.18/\sqrt{E}$ for electrons and $0.35/\sqrt{E}$
for hadrons, where $E$ is in \gev.  

The luminosity was measured \cite{lumi2} using the Bethe--Heitler reaction $ep
\rightarrow e\gamma p$ by a luminosity detector which consisted of two
independent systems: a lead--scintillator calorimeter
\cite{desy-92-066,*zfp:c63:391,*acpp:b32:2025} and a magnetic
spectrometer~\cite{nim:a565:572}.


\section{Theory}
\label{sec:theory}

The LO QCD processes relevant to the present measurements are the
direct and resolved photoproduction processes (Fig.~\ref{fig1}).
Higher-order processes include NLO diagrams and
fragmentation processes; a box-diagram term also contributes
significantly at next-to-next-to-leading order.

Two theoretical predictions are compared to the measurements presented
here. In the approach of FGH~\cite{epj:c21:303,epj:c34:191}, the LO
and NLO diagrams and the box-diagram term are calculated
explicitly. Fragmentation processes are also calculated in terms of a
fragmentation function in which a quark or gluon gives rise to a
photon; an experimentally determined non-perturbative parameterisation
is used as input to the theoretical calculation~\cite{epj:c19:89}.
The CTEQ6~\cite{jhep:0602:032} and AFG04~\cite{epj:c44:395} parton densities are used for the proton and photon
respectively; the use of alternatives altered the results by typically
5\%, which was small compared to the other uncertainties on the theory. 
The authors stress that their NLO calculation must include
fragmentation terms to give a well-defined result. Fragmentation and
box terms contribute each about 10\% to the total cross section.
Theoretical uncertainties arise due to the choice of renormalisation,
factorisation and fragmentation scales. They were estimated, using a
more conservative approach~\cite{priv:fgh:2013} than in the original published
paper~\cite{epj:c21:303}, by varying the renormalisation scale by
factors of 0.5 and 2.0, since this gave the largest effect on the
cross sections.

The $k_T$-factorisation method used by
LMZ~\cite{pr:d72:054002,pr:d81:094027,pr:d88:074001} makes use of
unintegrated parton densities in the proton, using the KMR
formalism~\cite{kmr} based on the MRST08 proton parton
densities~\cite{epj:c63:189}.  Fragmentation terms are not
included. The box diagram is included together with $2\to3$
subprocesses to represent the LO direct and resolved photon
contributions.  Uncertainties were evaluated as provided by LMZ.

All results are presented at the hadron level, and 
to make use of the predictions, cuts equivalent to the experimental
kinematic selections including the photon isolation (see Section
\ref{sec-selec}) were applied at the parton level.  Hadronisation
corrections were then evaluated (Section \ref{sec-mc}) and applied
to each of the calculations to enable the predictions to be compared
to the experimental data.


\section{Monte Carlo event simulation}
\label{sec-mc}
Monte Carlo (MC) event samples were generated to evaluate the detector
acceptance and event-reconstruction efficiency, and to provide signal
and background distributions.  The program \PYTHIA
6.416~\cite{jhep:0605:026} was used to generate the direct and
resolved prompt-photon processes, and also $2\to2$ parton-parton
scattering processes not involving photons (``dijet events'').  For
these purposes, CTEQ4~\cite{pr:d55:1280} and GRV~\cite{grv} 
parton densities were used.  The dijet event samples were generated to
enable background events to be extracted and used in the analysis.
Backgrounds to the isolated photons measured here arise from decays of
neutral mesons in hadronic jets where the decay products create an
energy cluster in the BCAL that passes the selection criteria for a
photon.  In \PYTHIA dijet events, a photon can also be radiated from
an incoming or outgoing quark.  Events in which a high-energy photon
was radiated from a quark or lepton (``radiative events'') were not
included in the final background samples but were defined, in
accordance with theory, as a component of the signal.

As a check and to enable systematic uncertainties to be estimated,
event samples were also generated using the \HERWIG 6.510
program~\cite{jhep:0101:010}. The cluster-based hadronisation scheme
of \HERWIG provides an alternative to the string-based scheme of
\PYTHIA.

The generated MC events were passed through the ZEUS detector and
trigger simulation programs based on {\sc Geant}
3.21~\cite{tech:cern-dd-ee-84-1}. They were then reconstructed and analysed
using the same programs as used for the data.  The hadronisation corrections to
the theory calculations were evaluated using \PYTHIA and
\HERWIG, and lowered the theoretical prediction by typically 
10\%. \PYTHIA and \HERWIG are in agreement to a few percent; 
\PYTHIA was used to provide the numbers for the present analysis.
No uncertainties were applied to these corrections.  They were
calculated by running the same jet algorithm and event selections,
including the isolation criterion, on the generated partons and on the
hadronised final state in the direct and resolved prompt-photon MC
events.


\section{Event selection and reconstruction}
\label{sec-selec}

 A three-level trigger system was used to select events online
\cite{zeus:1993:bluebook,uproc:chep:1992:222,nim:a580:1257}.
The first-level trigger required a loosely measured track in the CTD
and a minimum of energy deposited in the CAL. The event conditions
were tightened at the second level, and a high-energy photon candidate
was required at the third level.   Events were initially selected offline by requiring a
high-energy photon candidate of transverse energy $> 3.5 \gev$
recorded in the BCAL.  To reduce background from non-$ep$
collisions, events were required to have a reconstructed vertex
position, $Z_{\mathrm{vtx}}$, within the range $|Z_{\mathrm{vtx}}|<
40\,\mathrm{cm}$.  No scattered beam electron was permitted in the
detector, and photoproduction events were selected by the requirement
$0.2 < y_{\JB} < 0.7$, where $y_{\JB} = \sum \limits_i E_i(1-\cos
\theta_i)/2E_e$ and $E_e$ is the energy of the electron beam.  
Here, $E_i$ is the energy of the $i$-th CAL cell, $\theta_i$ is
its polar angle and the sum runs over all cells~\cite{pl:b303:183}.

Energy-flow objects
(EFOs)~\cite{epj:c1:81,*epj:c6:43,*briskin:phd:1998} were constructed
from clusters of calorimeter cells with signals, associated with
tracks when appropriate. Tracks not associated with calorimeter
clusters were also included.  Photon candidates were identified as
EFOs with no associated track, and with at least $90\%$ of the
reconstructed energy measured in the BEMC. Those EFOs with wider
electromagnetic showers than are typical for a single photon were
accepted to make possible the evaluation of backgrounds. Each event
was required to contain a photon candidate with a reconstructed
transverse energy, $E_T^{\gamma}$, in the range
\mbox{$6 <E_T^{\gamma}<15\,\mathrm{\gev\ }$} and with pseudorapidity,
$\eta^{\gamma}$, in the range $-0.7 < \eta^{\gamma} < 0.9$.

Jet reconstruction was performed, making use of all the EFOs in the
event including photon candidates, by means of the $k_T$ clustering
algorithm~\cite{np:b406:187} in the $E$-scheme in the longitudinally
invariant inclusive mode~\cite{pr:d48:3160} with the radius parameter
set to 1.0.  The jets were required to have transverse energy,
$E_T^{\mathrm{jet}}$, between 4 and 35\gev\ and to lie within the
pseudorapidity, $\eta^{\mathrm{jet}}$, range $-1.5
<\eta^{\mathrm{jet}} < 1.8$. By construction, one of the jets found by
this procedure corresponds to or includes the photon candidate. An
additional accompanying jet was required in the non-inclusive
measurements; if more than one was found, that with the highest
$E_T^{\mathrm{jet}}$ was used.  In this kinematic region, the
resolution of the jet transverse energy was about 15--20\%, estimated
using MC simulations.

To reduce the fragmentation contribution and the background from the
decay of neutral mesons within jets, the photon candidate was required
to be isolated from the reconstructed tracks and other hadronic
activity. High-\ET\ photons radiated from beam leptons were also
suppressed by requiring no observed scattered lepton in the apparatus.
The isolation from tracks was applied to exclude radiating electrons,
and was achieved by demanding $\Delta R>0.2$, where $\Delta R =
\sqrt{(\Delta \phi)^2 + (\Delta\eta)^2}$ is the distance to the
nearest reconstructed track with momentum greater than
$250\,\mathrm{MeV}$ in the $\eta \-- \phi$ plane, where $\phi$ is the
azimuthal angle. This condition was applied only at the detector
level, and not in the hadron- or parton-level calculations. Isolation
from other hadronic activity was imposed by requiring that the
photon-candidate EFO had at least $90 \%$ of the total energy of the
reconstructed jet of which it formed a part.  These selections gave
17441 events with an inclusive-photon candidate and 12450 events with
a photon candidate and an accompanying jet.

\section{Extraction of the photon signal}

The selected samples contain a large admixture of background
events in which one or more neutral mesons, such as $\pi^0$ and
$\eta$, decayed to photons, thereby producing a photon candidate in
the BEMC.  The photon signal was extracted statistically following the 
approach used in previous ZEUS analyses
\cite{pl:b595:86,epj:c54:371,pl:b687:16,pl:b715:88}.  

The photon signal was extracted from the background using the
energy-weighted width, measured in the $Z$ direction, of the BEMC
energy-cluster comprising the photon candidate.  This width was calculated
as $\langle\delta Z\rangle=
\sum \limits_i E_i|Z_i-Z_{\mathrm{cluster}}|$
{\large/}$( w_{\mathrm{cell}}\sum \limits_i E_i).$ Here, $Z_{i}$ is
the $Z$ position of the centre of the $i$-th cell,
$Z_{\mathrm{cluster}}$ is the energy-weighted centroid of the EFO
cluster, $w_{\mathrm{cell}}$ is the width of the cell in the $Z$
direction, and $E_i$ is the energy recorded in the cell. The sum runs
over all BEMC cells in the EFO.

\begin{figure}[p]
\vfill
\begin{center}
\epsfig{file=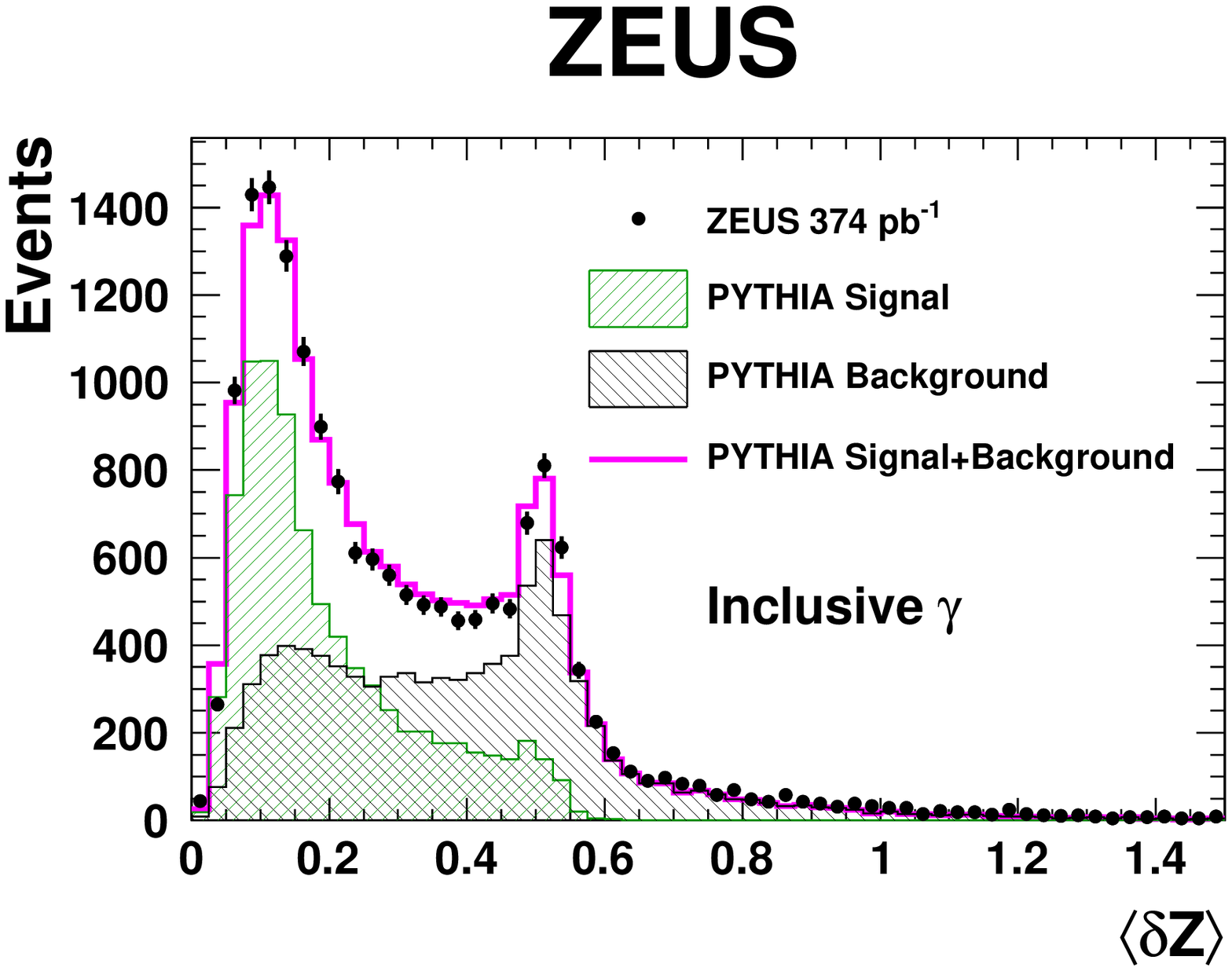,width=12cm}
\\[-70mm]
\hspace*{80mm}(a)\\[70mm]
\epsfig{file=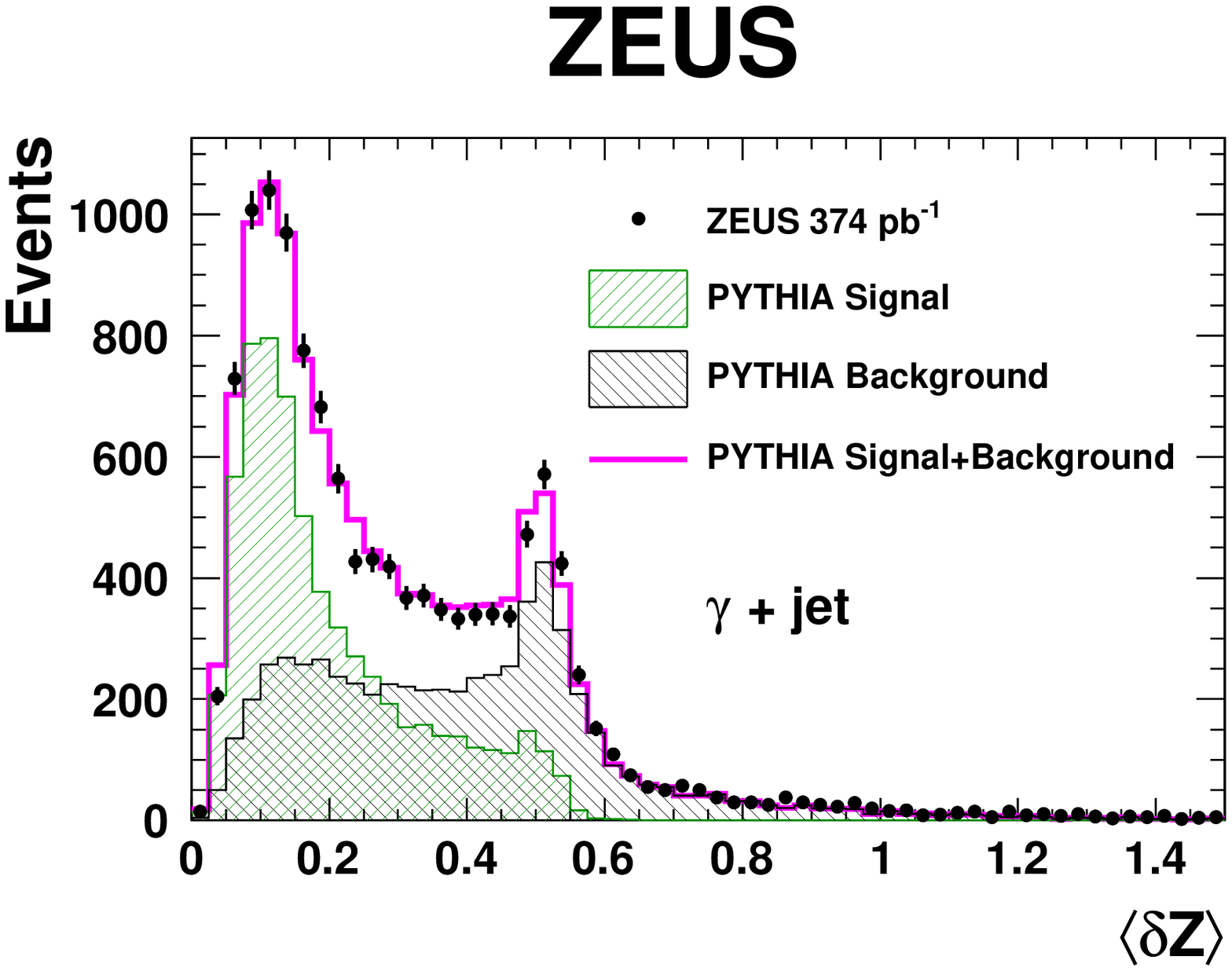,width=12cm}
\\[-70mm]
\hspace*{80mm}(b)\\[70mm]
\end{center}
\caption{\small 
 Distributions of $\langle \delta Z \rangle$ for (a) inclusive photon
  events, (b) events with a photon and an additional jet, showing the
  fitted signal and background components and their sum.  The error
  bars denote the statistical uncertainties on the data.  }
\label{fig:showers}
\vfill
\end{figure}

The global distribution of $\langle \delta Z \rangle$ in the data and
in the \PYTHIA\ MC are shown in Fig.~\ref{fig:showers} for inclusive photon
events and those containing an additional jet. The $\langle \delta Z \rangle$
distribution exhibits a double-peaked structure with the first peak at
$\approx 0.1$, associated with the photon signal, and the second peak at
$\approx 0.5$, dominated by the $\pi^0\rightarrow\gamma\gamma$
component of the background.

The number of isolated-photon events in the data is determined by a
$\chi^2$ fit to the $\langle \delta Z \rangle$ distribution in the
range $0.05<\langle \delta Z \rangle < 0.8$, varying the relative
fractions of the signal and background components as represented by
histogram templates obtained from the MC.  This is illustrated in
Fig.~\ref{fig:showers}, and a corresponding fit was performed for each
measured cross section bin, with $\chi^2$ values of typically 1.1 per
degree of freedom (i.e.~31/28). The extracted signals corresponded overall
to $8193\pm 156$ inclusive-photon events and $6262\pm 132$ events with
a photon and an accompanying jet.

\begin{figure}
\begin{center}
\epsfig{file=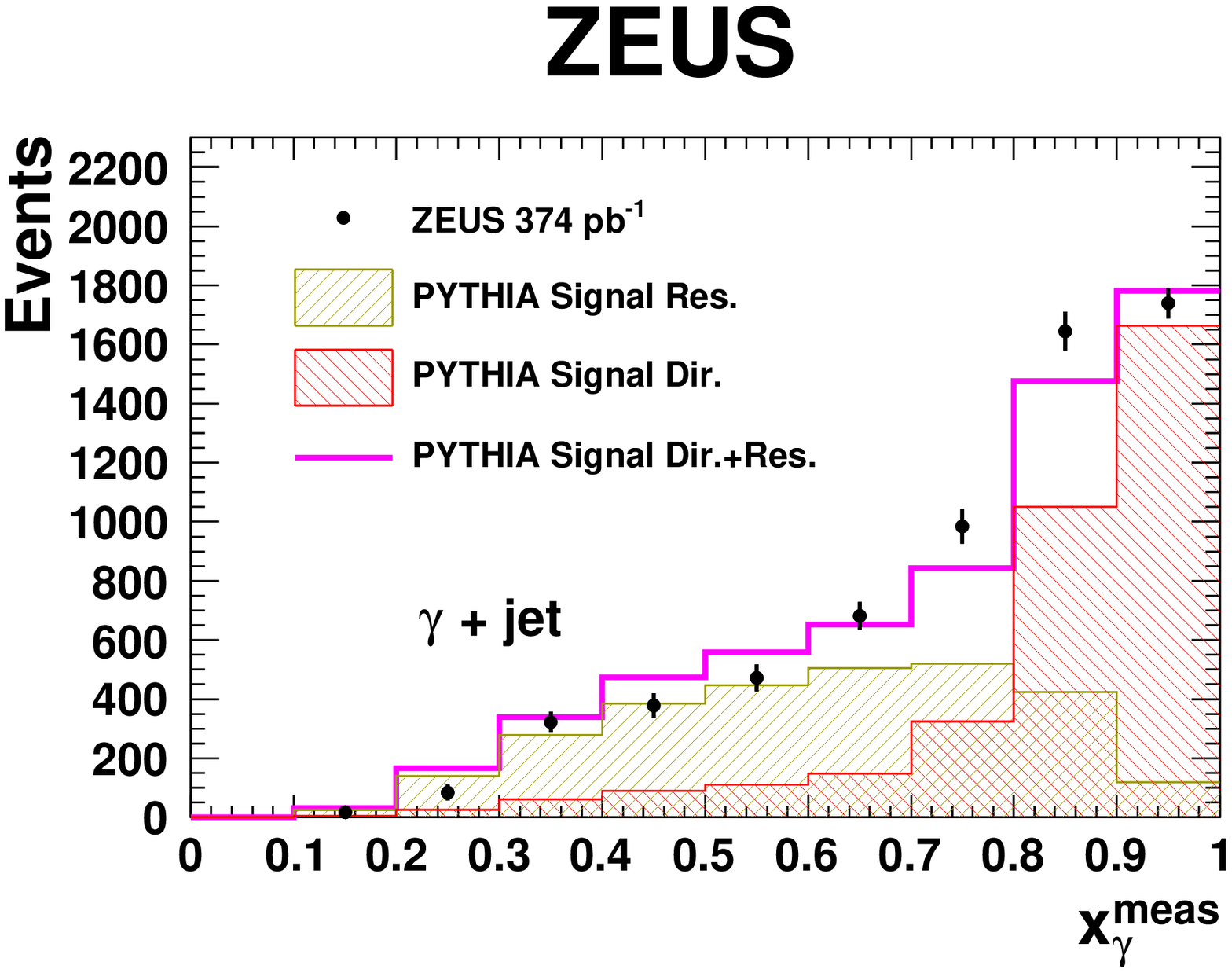,width=12cm}
\end{center}
\caption{\small  Events detected for different values of \xgamm, compared to
a mixture of {\sc Pythia}-generated direct and resolved events, using
the model described in the text. The simulated events were passed
through the detector simulation.  The kinematic ranges of the photons
and the jets are described in the text. No acceptance corrections were
applied at this stage.  }
\label{fig:xgamma}
\vfill
\end{figure}

A bin-by-bin correction method was used to determine the
production cross section, by means of the relationship

\begin{equation}
\frac{d\sigma}{dY} = \frac{\mathcal{A}\, N(\gamma)}{  
\mathcal{L} \, \Delta Y},
\end{equation}

where $N(\gamma)$ is the number of photons in a bin as extracted from the fit,
in events accompanied by a jet if required, and 
$\Delta Y$ is the bin width, $\mathcal{L}$ is the total integrated
luminosity, and $\mathcal{A}$ is the acceptance correction.  
The acceptance correction was calculated, using MC samples, as the
ratio of the number of events that were generated in the given bin to the number
that were obtained in the bin after event reconstruction.  Its value
was typically 1.2.  To evaluate the acceptances, allowance must be
made for the different acceptances for the direct and the resolved
processes, as modelled by
\PYTHIA.  These components can be substantially distinguished by means
of events containing a photon and a jet, in which the quantity
\begin{equation}
\xgamm =  \frac{\Egam+\Ejet-\pzgam-\pzjet}
{E^{\text{all}}- p_Z^{\text{all}}}.
\end{equation}
is a measure of the fraction of the incoming photon energy given to
the final-state photon and jet, at a lowest-order approximation.  The
energies and longitudinal momentum components of the photon
($\gamma$), the jet and all of the EFOs in the event were combined as
indicated.  Figure~\ref{fig:xgamma} shows the \xgamm\ distribution; a
peak close to unity is seen, which can be attributed to direct events,
and a tail at lower values due to resolved events. A reasonable
phenomenological description of the data can be obtained using a MC
sample consisting of a 50:40 mixture of \PYTHIA-simulated direct and
resolved events, as normalised to the data, with a 10\% admixture of
radiative events divided equally between direct and resolved. The
acceptance factors were calculated using this model.  Acceptance
factors calculated in this way were applied both to the inclusive and
to the jet data.  

The trigger efficiency was approximately flat above a photon
transverse energy of 4.5\gev, with a value of $87\pm2\%$.  This
includes a correction of 3.6\% which was applied to the trigger
acceptance modelled by the MC.  The correction was evaluated using DIS
samples, in data and MC, in which events with prompt photons were
triggered in an independent way.

A correction of typically 2\% was applied to subtract a contamination
of the sample by DIS events, which was determined using MC-simulated
DIS samples~\cite{pl:b715:88}.

\section{Systematic uncertainties}
\label{sec:syst}

The most significant sources of systematic uncertainty were evaluated
as follows:

\begin{itemize}

\item to allow for uncertainties in the simulation of the hadronic final 
state, the cross sections were recalculated using \HERWIG to model the
signal and background events.  The ensuing changes in the results
correspond to an uncertainty of typically up to 8\%, but rising to
18\% in the highest bin of \xgamm;

\item the energy of the photon candidate was varied by $\pm 2\%$ in the 
MC at the detector level. This value was obtained from a study of
energy-momentum conservation in Deeply Virtual Compton Scattering
events measured in the ZEUS detector, in which the final state
consisted of a photon and a scattered electron.  Independently, the
energy of the accompanying jet, when measured, was varied by an amount
decreasing from $\pm4.5\%$ to $\pm2.5\%$ as
\ETjet\ increases from 4\gev\ to above 10\gev. 
These values were obtained as described in a previous ZEUS
publication~\cite{pl:b715:88}.  Each of these contributions gave
variations in the measured cross sections of typically 5\%.
\end{itemize}

Further systematic uncertainties were evaluated as follows:

\begin{itemize}

\item the uncertainty in the acceptance due to the estimation of the 
relative fractions of direct and resolved events and radiative events
in the MC sample was estimated by varying these fractions by $\pm
15\%$ and $\pm 5\%$ respectively in absolute terms; the changes in the
cross sections were typically $\pm2\%$ in each case;

\item  the dependence of the result on the modelling of the hadronic 
background by the MC was investigated by varying the upper limit for
the $\langle
\delta Z\rangle$ fit in the range $[0.6, 1.0]$; this gave a $\pm
2\%$ variation.

\end{itemize}

Other sources of systematic uncertainty were found to be negligible
and were ignored. These included the modelling of the track-isolation
cut, the track-momentum cut, and the cuts on photon isolation, the
electromagnetic fraction of the photon shower, $y_{\JB}$ and
$Z_{\text{vtx}}$.  Except for the uncertainty on the modelling of the
hadronic final state, the major uncertainties were treated as
symmetric, and all the uncertainties were combined in quadrature.  The
common uncertainties of 2.0\% on the trigger efficiency and 1.9\% on
the luminosity measurement were not included in the tables and figures.

\section{Results}

\label{sec:results}

Differential cross sections were measured for the production of an
isolated photon inclusively, and with at least one accompanying jet,
in the kinematic region defined by $Q^2< 1 \, \mathrm{\gev}^2$, $0.2 <
y < 0.7,$ $-0.7< \eta^\gamma < 0.9$, $6 < E_T^{\gamma}<
15\,\mathrm{\gev} $, and where relevant $4 < \ETjet < 35$\gev\ and
$-1.5 <\etajet< 1.8$.  All quantities were evaluated at the hadron
level in the laboratory frame. Again, the jets were formed
according to the $k_T$ clustering algorithm with the radius parameter
set to 1.0.  Photon isolation was imposed such that at least $90 \%$
of the energy of the jet-like object containing the photon originated
from the photon.  If more than one accompanying jet was found within
the designated \etajet range in an event, that with highest \ETjet was
taken.  The integrated luminosity was $374\pm7\,\mathrm{pb}^{-1}$.

\begin{figure}

\vfill
\begin{center}
\epsfig{file=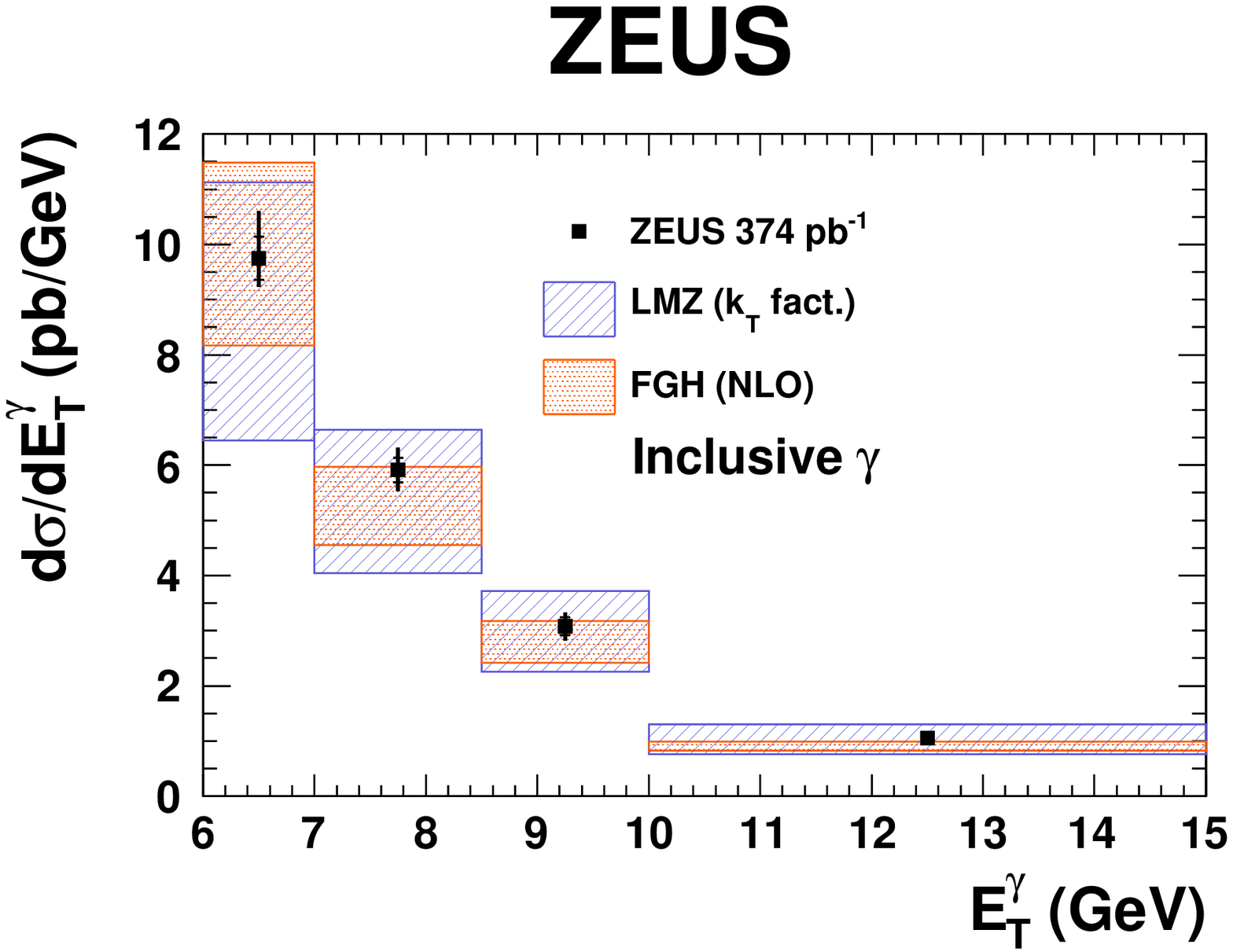,width=12cm}
\\[-70mm]
\hspace*{80mm}(a)\\[70mm]
\epsfig{file=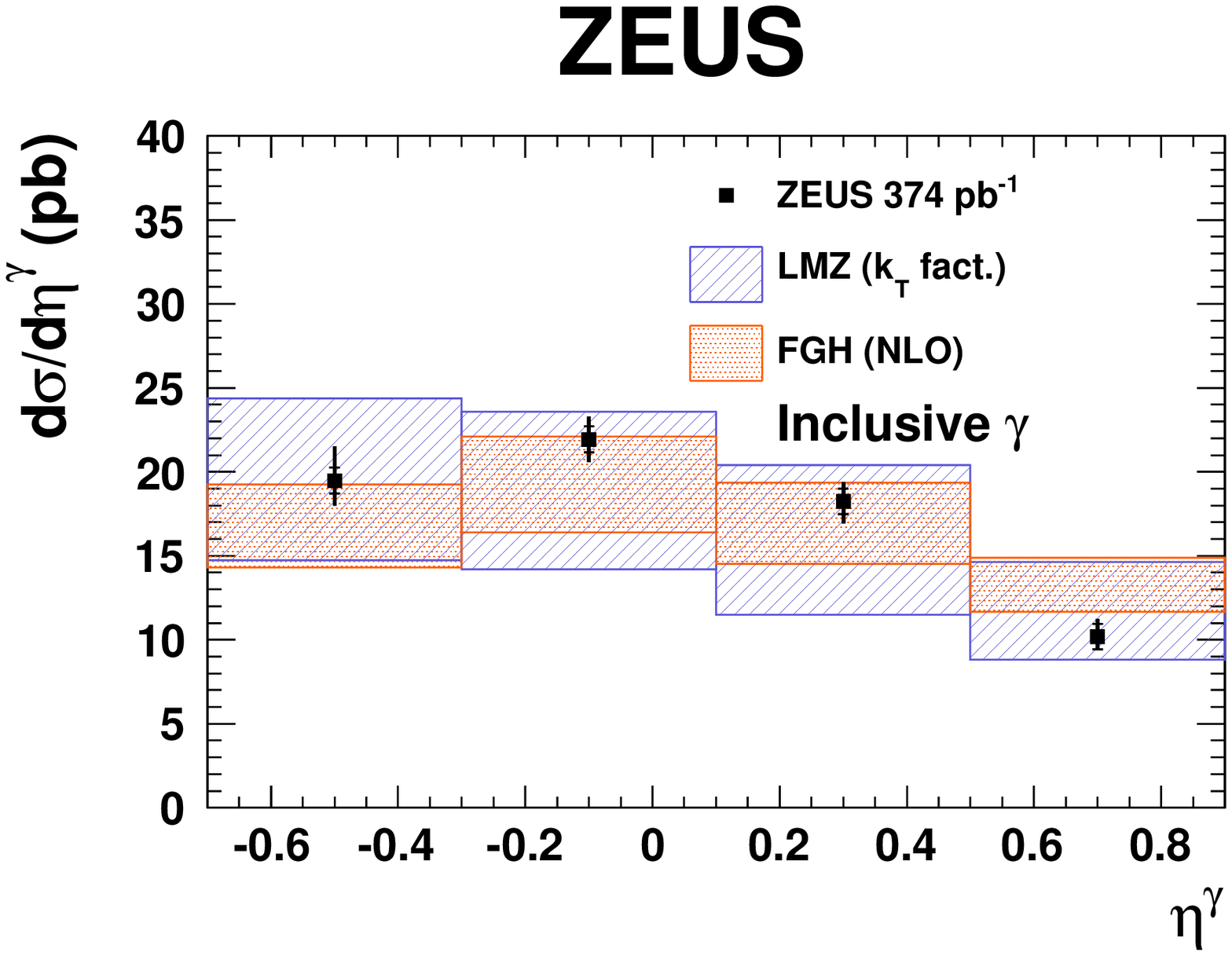,width=12cm}
\\[-70mm]
\hspace*{80mm}(b)\\[70mm]
\end{center}
\caption{\small  Differential cross sections as functions of (a) \ETgam and (b) \etagam  
for events containing an isolated photon, compared to predictions from
FGH and LMZ.  The kinematic region of the measurement is described in
the text.  The inner and outer error bars respectively denote
statistical uncertainties and statistical uncertainties combined with
systematic uncertainties in quadrature. The theoretical uncertainties are shown as
hatched and dotted bands.}
\label{fig:incgam}
\end{figure}

\begin{figure}

\vfill
\begin{center}
\epsfig{file=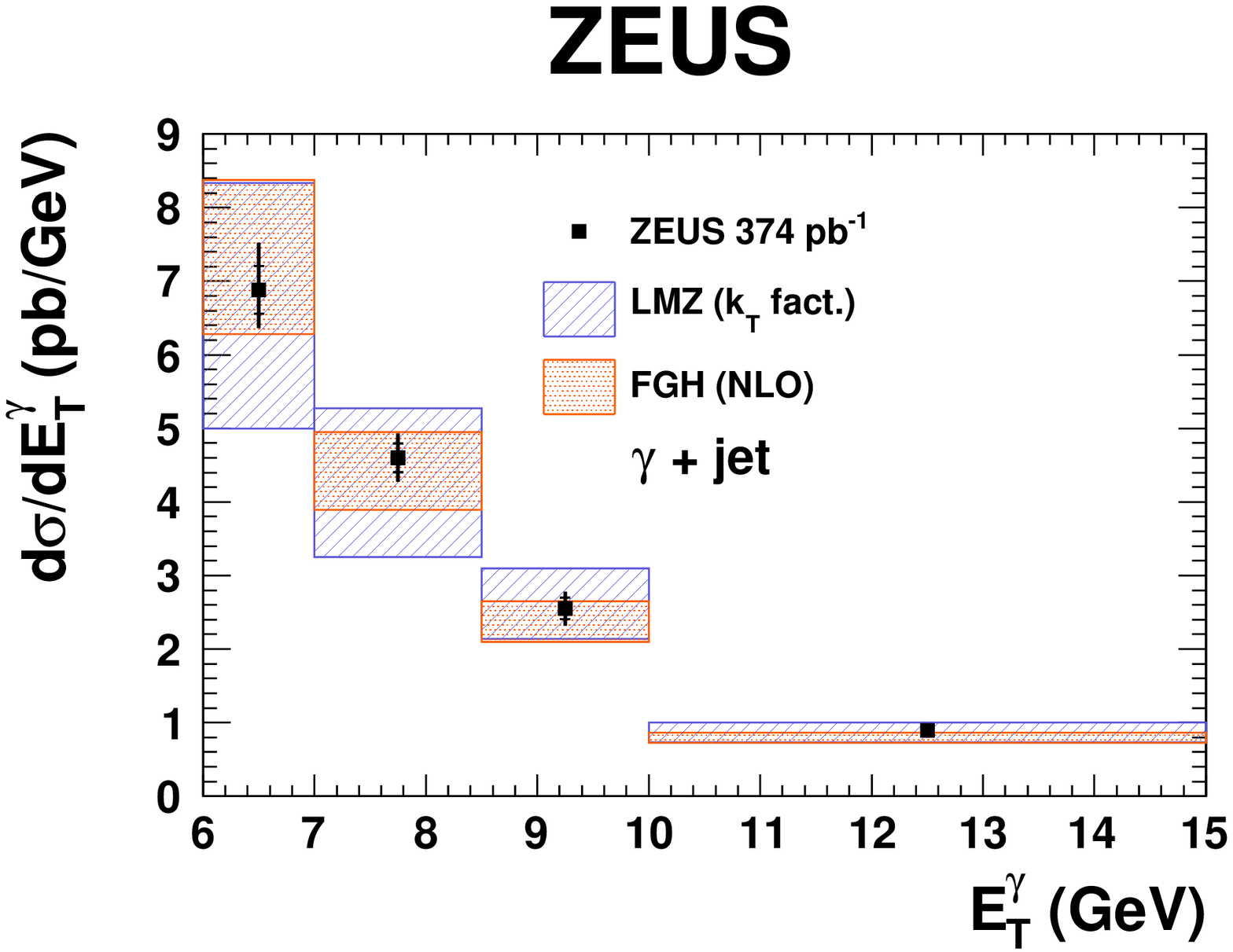,width=12cm}
\\[-70mm]
\hspace*{80mm}(a)\\[70mm]
\epsfig{file=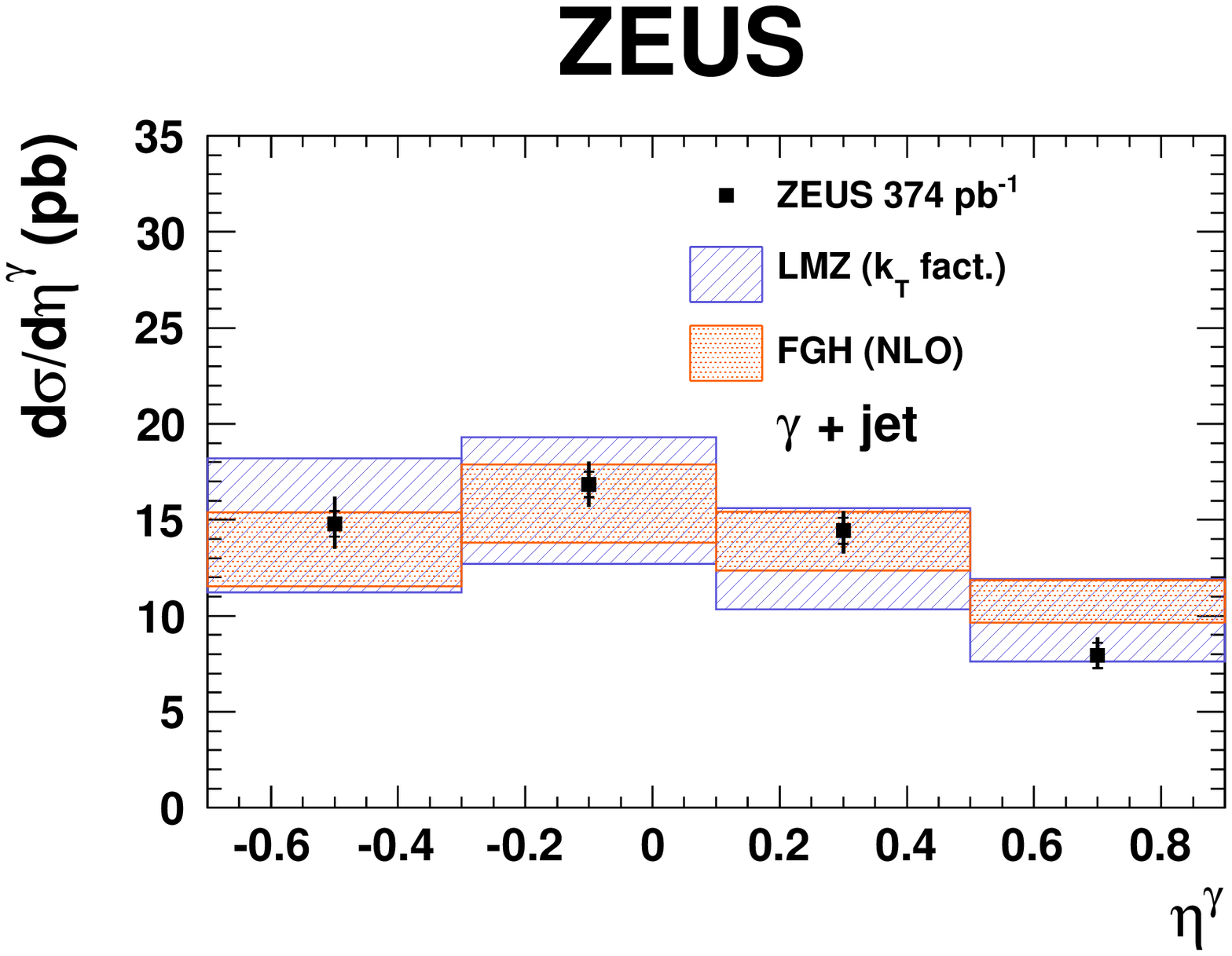,width=12cm}
\\[-70mm]
\hspace*{80mm}(b)\\[70mm]
\end{center}
\caption{\small  Differential cross sections as functions of (a) \ETgam and (b) \etagam, 
for events containing an isolated photon accompanied by a jet, 
compared to predictions from FGH and LMZ. 
  The kinematic region of the measurement is described in
the text.  The inner and outer error bars respectively denote
statistical uncertainties and statistical uncertainties combined with
systematic uncertainties in quadrature. The theoretical uncertainties are shown as
hatched and dotted bands.
}
\label{fig:jetgam}
\end{figure}

\begin{figure}

\vfill
\begin{center}
\epsfig{file=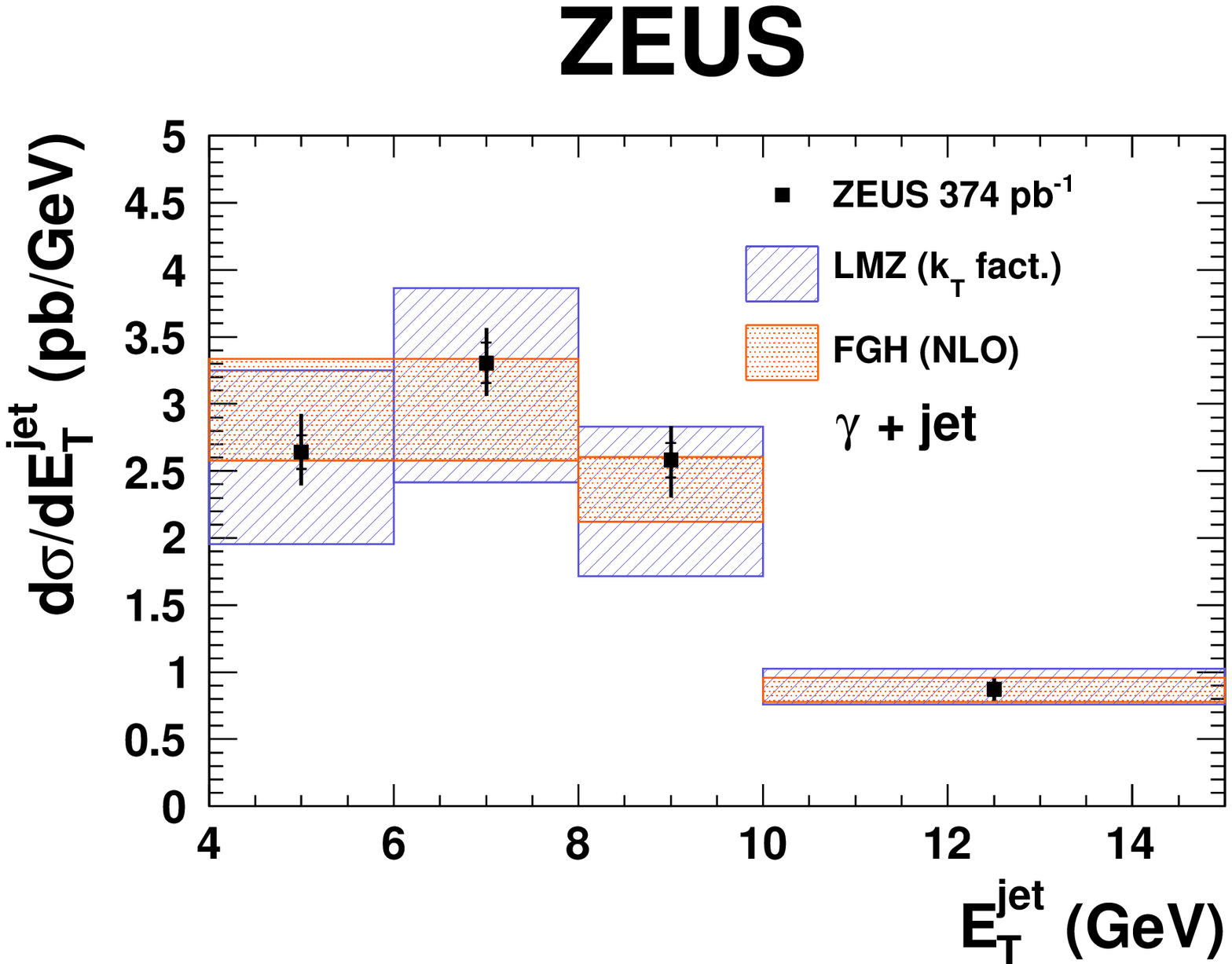,width=12cm}
\\[-70mm]
(a)\hspace*{65mm}~\\[70mm]
\epsfig{file=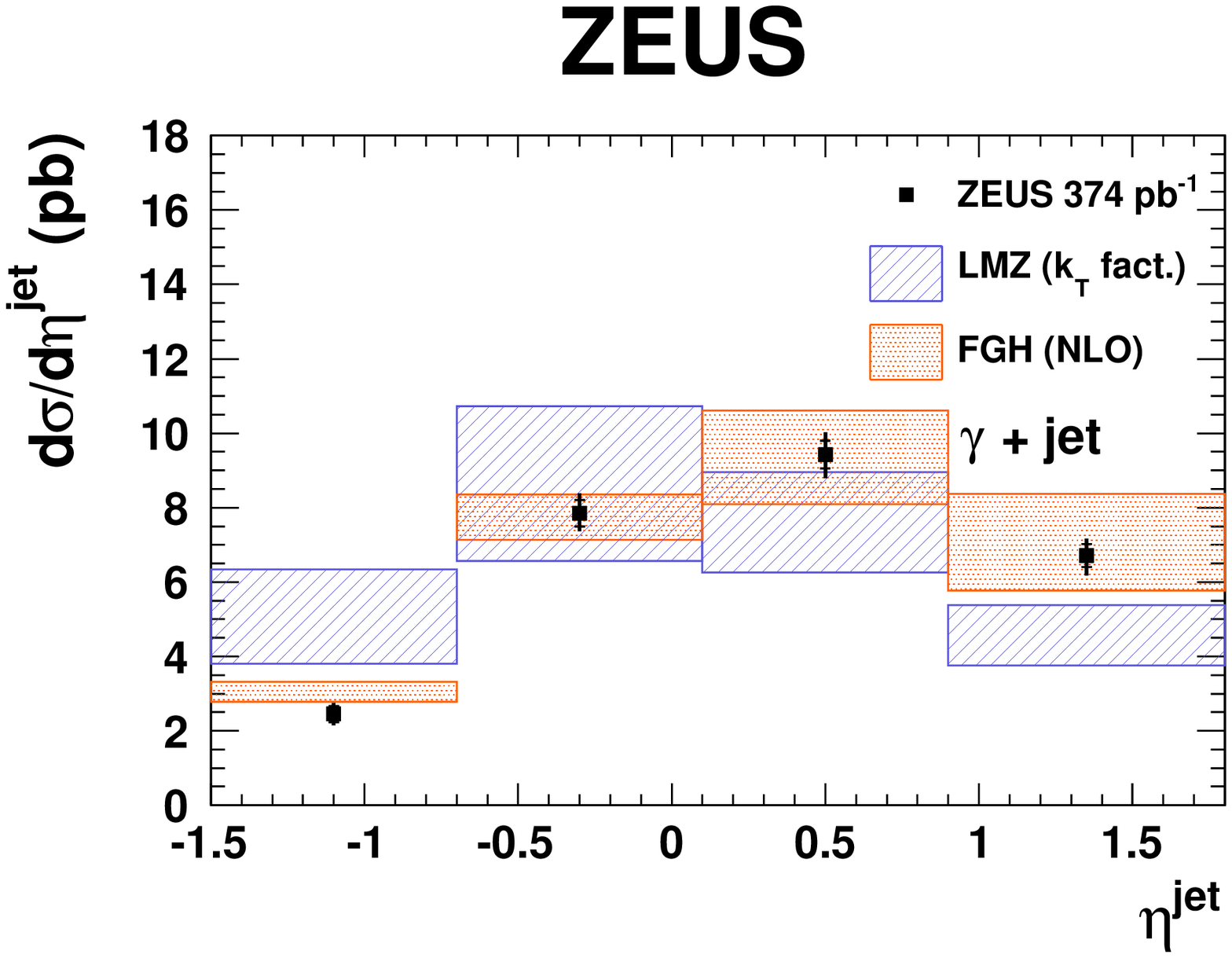,width=12cm}
\\[-70mm]
(b)\hspace*{65mm}~\\[70mm]
\end{center}
\caption{\small  Differential cross sections as functions of (a) \ETjet and (b) \etajet, 
for events containing an isolated photon accompanied by a jet,
compared to predictions from FGH and LMZ.  The kinematic region of the
measurement is described in the text.  The inner and outer error bars
respectively denote statistical uncertainties and statistical
uncertainties combined with systematic uncertainties in
quadrature. The theoretical uncertainties are shown as hatched and
dotted bands.  The first two FGH points in (a) have been averaged into
a single bin for calculational reasons.  }
\label{fig:jet}
\end{figure}

\begin{figure}

\vfill
\begin{center}
\epsfig{file=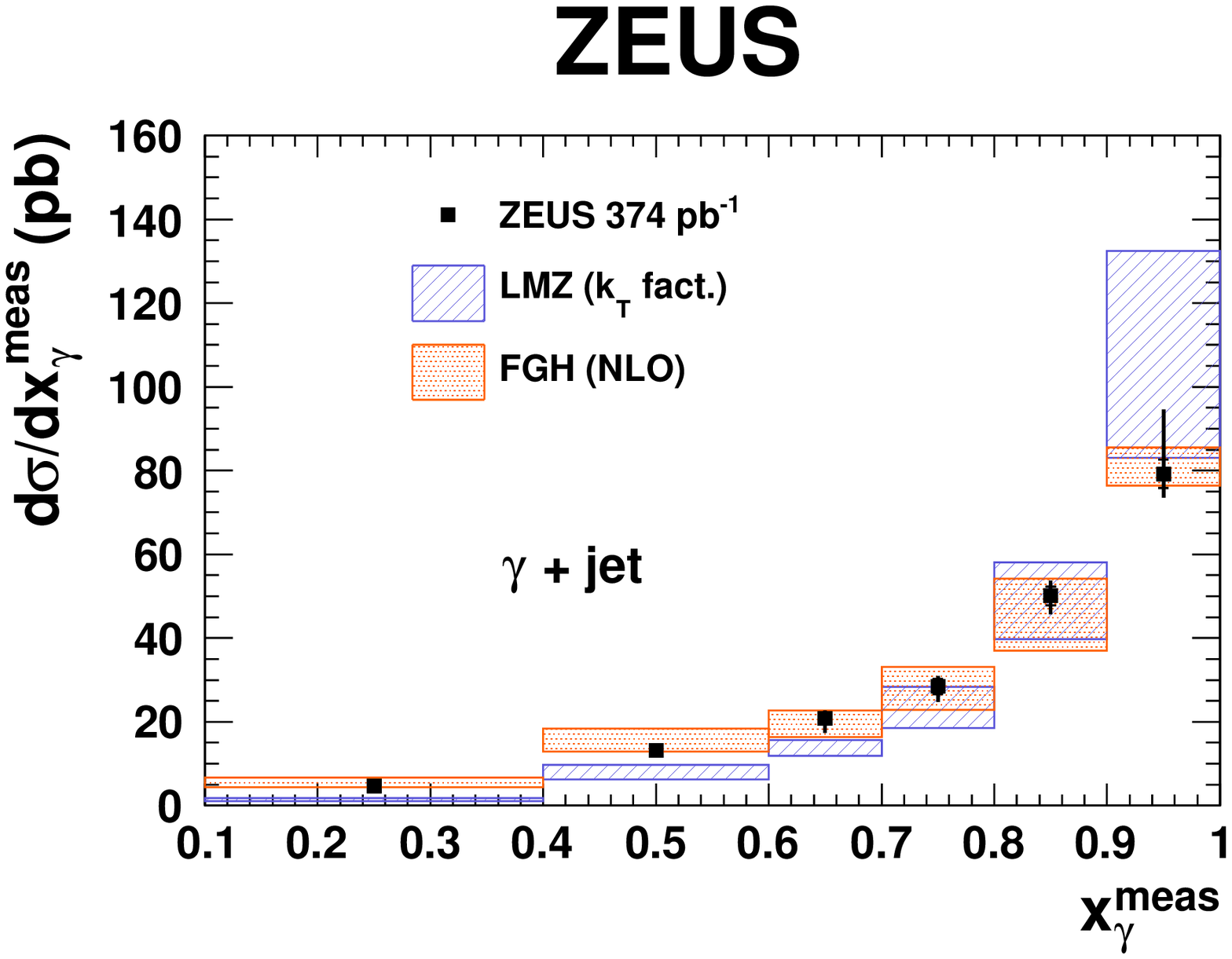,width=12cm}
\end{center}
\caption{\small  Differential cross section as a function of \xgamm, 
for events containing an isolated photon and a jet, compared to
predictions from FGH and LMZ.  
 The kinematic region of the measurement is described in
the text.  The inner and outer error bars respectively denote
statistical uncertainties and statistical uncertainties combined with
systematic uncertainties in quadrature. The theoretical uncertainties are shown as
hatched and dotted bands.
}
\label{fig:xgammb}
\end{figure}

The differential cross sections as functions of $E_T^{\gamma}$,
$\eta^{\gamma}$, $\ETjet$, $\etajet$ and \xgamm are shown in
Figs.~\ref{fig:incgam}--\ref{fig:jet}, and \ref{fig:xgammb}, and given
in Tables \ref{tab:eti}--\ref{tab:xg}. Cross sections in $\ETjet$
above 15\gev\ are omitted from Table 5 and Fig.~\ref{fig:jet}(a) owing
to limited statistics, but this kinematic region is included in the
other cross-section measurements.  The theoretical predictions
described in Section~\ref{sec:theory} are compared to the
measurements; theoretical uncertainties are indicated by the width of
the respective shaded areas. The NLO-based predictions from FGH
describe the distributions well. The predictions of LMZ, within their
uncertainties, also describe the photon distributions well, but give a
less good description at low $\etajet$ and low $\xgamm$.  The
experimental uncertainties are substantially smaller than those of the
theory.

\section{Conclusions}

The production of inclusive isolated photons and photons with an
accompanying jet has been measured in photoproduction with the ZEUS
detector at HERA using an integrated luminosity of
$374\,\pm7\,\mathrm{pb}^{-1}$.  The present results improve on earlier
ZEUS results, which were made with lower integrated luminosities.
Differential cross sections are presented as functions of the
transverse energy and the pseudorapidity of the photon and the jet,
and \xgamm, where the kinematic region is defined in the
laboratory frame by: $Q^2< 1 \,
\mathrm{\Gev}^2$, $0.2 < y < 0.7$,
$-0.7< \eta^\gamma < 0.9$, $6 < E_T^{\gamma}< 15\,\mathrm{\Gev\ } $
and, where a jet is required, $4< \ETjet <35$\gev\ and $-1.5 <\etajet<
1.8$.  Photon isolation was imposed such that at least $90 \%$ of the
energy of the jet-like object containing the photon originated from
the photon. The NLO-based predictions of Fontannaz, Guillet and
Heinrich reproduce the measured experimental distributions well.  The
$k_T$-factorisation approach of Lipatov, Malyshev and Zotov describes
the photon distributions well but gives a less good description of the
jet-based variables.

\section*{Acknowledgements}
\label{sec-ack}

\Zacknowledge\ We also thank M. Fontannaz, G. Heinrich, A. Lipatov, M. Malyshev and N. Zotov for providing assistance and theoretical results.

\vfill\eject

\include{prp2013a-ref}
\raggedright{
\providecommand{\etal}{et al.\xspace}
\providecommand{\coll}{Collaboration}
\catcode`\@=11
\def\@bibitem#1{%
\ifmc@bstsupport
  \mc@iftail{#1}%
    {;\newline\ignorespaces}%
    {\ifmc@first\else.\fi\orig@bibitem{#1}}
  \mc@firstfalse
\else
  \mc@iftail{#1}%
    {\ignorespaces}%
    {\orig@bibitem{#1}}%
\fi}%
\catcode`\@=12


}
\begin{table}
\begin{center}
\begin{tabular}{|rcr|c|c|}
\hline
\multicolumn{3}{|c|}{$E_T^{\gamma}$ range}   &   & had.\ corr.     \\[-1.0ex]
\multicolumn{3}{|c|}{ (GeV)}   &  \raisebox{1.5ex}{$\frac{d\sigma}{dE^{\gamma}_T}$ ($\mathrm{pb}\,\mathrm{GeV}^{-1})$}   &    \\
\hline
6 & -- & 7   & $9.75 \pm  0.39\,\mathrm{(stat.)}\,^{+0.75}_{-0.35}\,\mathrm{(syst.)}$ &0.88 \\
7 & -- & 8.5 & $5.91 \pm  0.22\,\mathrm{(stat.)}\,^{+0.33}_{-0.31}\,\mathrm{(syst.)}$ &0.90 \\
8.5 & -- & 10  & $3.08 \pm  0.16\,\mathrm{(stat.)}\,^{+0.20}_{-0.20}\,\mathrm{(syst.)}$ &0.93 \\
10 & -- & 15 & $1.06 \pm  0.05\,\mathrm{(stat.)}\,^{+0.06}_{-0.09}\,\mathrm{(syst.)}$ &0.96 \\
\hline
\end{tabular}
\end{center}
\caption{Measured differential cross-section $\frac{d\sigma}{dE^{\gamma}_T}$ for inclusive photons. 
The multiplicative hadronisation correction  applied to the theory is given under ``had.\ corr.''.\label{tab:eti}}
\end{table}

\begin{table}
\begin{center}
\begin{tabular}{|rcr|c|c|}
\hline
\multicolumn{3}{|c|}{ $\eta^{\gamma}$ range }  & \multicolumn{1}{|c|}{ $\frac{d\sigma}{d\eta^{\gamma}}$ ($\mathrm{pb}$)}     & had.\ corr. \\[0.5mm]
\hline
--0.7 & -- & --0.3   & $19.48 \pm  0.77\,\mathrm{(stat.)}\,^{+1.91}_{-1.27}\,\mathrm{(syst.)}$ &0.94 \\
--0.3 & -- & 0.1   & $21.94 \pm  0.76\,\mathrm{(stat.)}\,^{+1.12}_{-1.12}\,\mathrm{(syst.)}$ &0.92 \\
0.1 & -- & 0.5  & $18.24 \pm  0.76\,\mathrm{(stat.)}\,^{+0.87}_{-1.07}\,\mathrm{(syst.)}$ &0.89 \\
0.5 & -- & 0.9 & $10.19 \pm  0.75\,\mathrm{(stat.)}\,^{+0.76}_{-0.20}\,\mathrm{(syst.)}$ &0.88 \\
\hline
\end{tabular}
\end{center}
\caption{Measured differential cross-section $\frac{d\sigma}{d\eta^{\gamma}}$ for inclusive photons, and hadronisation correction.\label{tab:etai}}
\end{table}

\begin{table}
\begin{center}
\begin{tabular}{|rcr|c|c|}
\hline
\multicolumn{3}{|c|}{$E_T^{\gamma}$ range}   &       & had.\ corr. \\[-1.0ex]
\multicolumn{3}{|c|}{ (GeV)}   &  \raisebox{1.5ex}{$\frac{d\sigma}{dE^{\gamma}_T}$ ($\mathrm{pb}\,\mathrm{GeV}^{-1})$} &     \\
\hline
6 & -- & 7   & $6.88 \pm  0.33\,\mathrm{(stat.)}\,^{+0.55}_{-0.41}\,\mathrm{(syst.)}$ &0.83 \\
7 & -- & 8.5 & $4.60 \pm  0.19\,\mathrm{(stat.)}\,^{+0.28}_{-0.25}\,\mathrm{(syst.)}$ &0.87 \\
8.5 & -- & 10  & $2.55 \pm  0.14\,\mathrm{(stat.)}\,^{+0.17}_{-0.19}\,\mathrm{(syst.)}$ &0.90 \\
10 & -- & 15 & $0.90 \pm  0.04\,\mathrm{(stat.)}\,^{+0.05}_{-0.07}\,\mathrm{(syst.)}$ &0.93 \\
\hline
\end{tabular}
\end{center}
\caption{Measured differential cross-section $\frac{d\sigma}{dE^{\gamma}_T}$ for photons accompanied by a jet, and hadronisation correction. \label{tab:etg}}
\end{table}

\begin{table}
\begin{center}
\begin{tabular}{|rcr|c|c|}
\hline
\multicolumn{3}{|c|}{ $\eta^{\gamma}$ range }  & \multicolumn{1}{|c|}{ $\frac{d\sigma}{d\eta^{\gamma}}$ ($\mathrm{pb}$)}     & had.\ corr. \\[0.5mm]
\hline
--0.7 & -- & --0.3   & $14.80 \pm  0.66\,\mathrm{(stat.)}\,^{+1.24}_{-1.14}\,\mathrm{(syst.)}$ &0.90 \\
--0.3 & -- & 0.1   & $16.86 \pm  0.66\,\mathrm{(stat.)}\,^{+0.97}_{-0.97}\,\mathrm{(syst.)}$ &0.88 \\
0.1 & -- & 0.5  & $14.43 \pm  0.67\,\mathrm{(stat.)}\,^{+0.75}_{-0.97}\,\mathrm{(syst.)}$ &0.86 \\
0.5 & -- & 0.9 & $7.95 \pm  0.66\,\mathrm{(stat.)}\,^{+0.67}_{-0.23}\,\mathrm{(syst.)}$ &0.85 \\
\hline
\end{tabular}
\end{center}
\caption{Measured differential cross-section $\frac{d\sigma}{d\eta^{\gamma}}$ for photons accompanied by a jet, and hadronisation correction.\label{tab:etag}}
\end{table}

\begin{table}
\begin{center}
\begin{tabular}{|rcr|c|c|}
\hline
\multicolumn{3}{|c|}{$E_T^{\jet}$ range}   &       & had.\ corr. \\[-1.0ex]
\multicolumn{3}{|c|}{ (GeV)}   &  \raisebox{1.5ex}{$\frac{d\sigma}{dE^{\jet}_T}$ ($\mathrm{pb}\,\mathrm{GeV}^{-1})$}   &   \\
\hline
4 & -- & 6   & $2.64 \pm  0.13\,\mathrm{(stat.)}\,^{+0.26}_{-0.21}\,\mathrm{(syst.)}$ &0.86 \\
6 & -- & 8   & $3.31 \pm  0.15\,\mathrm{(stat.)}\,^{+0.21}_{-0.19}\,\mathrm{(syst.)}$ &0.79 \\
8  & -- & 10  & $2.58 \pm  0.13\,\mathrm{(stat.)}\,^{+0.22}_{-0.24}\,\mathrm{(syst.)}$ &0.90 \\
10 & -- & 15 & $0.87 \pm  0.05\,\mathrm{(stat.)}\,^{+0.07}_{-0.07}\,\mathrm{(syst.)}$ &0.98 \\
\hline
\end{tabular}
\end{center}
\caption{Measured differential cross-section
$\frac{d\sigma}{dE^{\jet}_T}$ for photons accompanied by a jet, and
hadronisation correction.
\label{tab:etj}}
\end{table}

\begin{table}
\begin{center}
\begin{tabular}{|rcr|c|c|}
\hline
\multicolumn{3}{|c|}{ $\eta^{\jet}$ range }  & \multicolumn{1}{|c|}{ $\frac{d\sigma}{d\eta^{\jet}}$ ($\mathrm{pb}$)}     & had.\ corr. \\[0.5mm]
\hline
--1.5 & -- & --0.7   & $2.46 \pm  0.22\,\mathrm{(stat.)}\,^{+0.21}_{-0.22}\,\mathrm{(syst.)}$ &0.71 \\
--0.7 & -- & 0.1   & $7.85 \pm  0.36\,\mathrm{(stat.)}\,^{+0.39}_{-0.31}\,\mathrm{(syst.)}$ &0.80 \\
0.1 & -- & 0.9  & $9.42 \pm  0.37\,\mathrm{(stat.)}\,^{+0.47}_{-0.51}\,\mathrm{(syst.)}$ &0.96 \\
0.9 & -- & 1.8 & $6.71 \pm  0.31\,\mathrm{(stat.)}\,^{+0.34}_{-0.43}\,\mathrm{(syst.)}$ &1.11 \\
\hline
\end{tabular}
\end{center}
\caption{Measured differential cross-section
$\frac{d\sigma}{d\eta^{\jet}}$ for photons accompanied by a jet, and
hadronisation correction.
\label{tab:etaj}}
\end{table}

\begin{table}
\begin{center}
\begin{tabular}{|rcr|c|c|}
\hline
\multicolumn{3}{|c|}{ $\xgamm$ range }  & \multicolumn{1}{|c|}{ $\frac{d\sigma}{d\xgamm}$ ($\mathrm{pb}$)} & had.\ corr. \\[1.0mm]
\hline
  0.1 & -- &  0.4   & $4.66 \pm  0.54\,\mathrm{(stat.)}\,^{+0.40}_{-0.41}\,\mathrm{(syst.)}$ &0.67 \\
  0.4 & -- &   0.6   & $13.18 \pm  1.07\,\mathrm{(stat.)}\,^{+0.95}_{-1.05}\,\mathrm{(syst.)}$ &0.88 \\
  0.6 & -- & 0.7   & $20.77 \pm  1.62\,\mathrm{(stat.)}\,^{+1.05}_{-3.06}\,\mathrm{(syst.)}$ &0.98 \\
  0.7 & -- & 0.8   & $28.42 \pm  1.83\,\mathrm{(stat.)}\,^{+1.76}_{-3.13}\,\mathrm{(syst.)}$ &1.32 \\
 0.8 & -- & 0.9  & $50.07 \pm  2.30\,\mathrm{(stat.)}\,^{+2.92}_{-3.81}\,\mathrm{(syst.)}$ &1.72 \\
0.9 & -- & 1.0  & $79.23 \pm  3.41\,\mathrm{(stat.)}\,^{+14.95}_{-4.62}\,\mathrm{(syst.)}$ &0.68 \\
\hline
\end{tabular}
\end{center}
\caption{Measured differential cross-section $\frac{d\sigma}{d\xgamm}$
for photons accompanied by a jet, and hadronisation correction.
\label{tab:xg}}
\end{table}

\newpage
\clearpage
%

%

%
\end{document}